\documentclass[%
  aps,
  pre,
  reprint,
  superscriptaddress,
  floatfix,
  nofootinbib,
  amsmath,amssymb,
  longbibliography,
]{revtex4-2}

\usepackage{graphicx}
\graphicspath{{../figures/}{figures/}}
\usepackage{mathtools}
\usepackage{bm}
\usepackage{booktabs}
\usepackage{array}
\usepackage{tabularx}
\usepackage{multirow}
\usepackage{placeins}
\usepackage{hyperref}
\usepackage{xcolor}
\hypersetup{
  colorlinks=true,
  linkcolor=blue!50!black,
  citecolor=blue!50!black,
  urlcolor=blue!60!black,
}

\newcommand{\Targmax}{T_{\mathrm{argmax}}}

\newcommand{\Nh}{N_{h}}
\newcommand{\Tlam}{L_{\mathrm{lam}}}
\newcommand{\Tpl}{T_{\mathrm{pl}}}
\newcommand{\Ph}{P(h)}
\newcommand{\Pdh}{P(\delta\mid h)}

\newcommand{\Levy}{L\'{e}vy}
\newcommand{\BIB}{BIB}
\newcommand{\BO}{BO}

\begin{document}

\title{Internal-state criticality in Bayesian--inverse-Bayesian
inference}

\author{Kazuto Sasai}
\email{kazuto.sasai.z@vc.ibaraki.ac.jp}
\affiliation{Faculty of Applied Science and Engineering,
Ibaraki University, Hitachi, Ibaraki 316-8511, Japan}
\author{Yukio-Pegio Gunji}
\email{yukio@waseda.jp}
\affiliation{Department of Intermedia Art and Science, School of Fundamental Science and Engineering, Waseda University, Tokyo 169-8555, Japan}
\date{\today}

\begin{abstract}
We propose Bayesian--inverse-Bayesian (\BIB) inference in repeated
games as a minimal model linking Bayesian inference, statistical
mechanics, and heavy-tailed (power-law) statistics, and as a broadly
applicable testbed for the dynamics of inference.  As a concrete
instantiation we simulate repeated $N$-hand cyclic-dominance
rock-paper-scissors (RPS), a discrete setting in which Nash-targeting
algorithms collapse to uniform random play, so that any non-trivial
dynamics must originate internally.
Across a multi-axis sweep of design, window, and pair conditions, the
\BIB{} dynamics remain in the same internal critical state across
implementation designs and opponent classes, the argmax-persistence
distribution staying a heavy-tailed power law with exponent
$\alpha\approx 1.43$ at the canonical window ($m=50$, $N=3$).  Along the
window and alphabet axes the exponent is not constant but drifts
systematically toward the universal $3/2$ as the finite-sample residual
$(N-1)/(2m)$ vanishes.  Bayes-only (\BO) inference, which lacks the inverse step, shows no
analogous universality.  Under the same model selection its
persistence-time distributions are not power laws, so it has no \Levy{}
exponent.  The \BIB{} laminar-phase length distribution follows a
truncated power law close to the universal $\alpha=3/2$ of on-off
intermittency.  Because both observables are first-passage reads of one
driftless log-posterior walk, what is robust across conditions is the
critical, zero-drift state itself, evidenced by the cross-design data
collapse rather than by any particular exponent value.  The
same exponent persists, invariant, across the
hypothesis count $\Nh$ within a finite-size-scaling core regime,
whereas the \BO{} exponent varies widely over the $\Nh$ sweep.  The
cutoff time and the posterior spread both obey finite-size scaling in
$\Nh$, the latter close to its information-theoretic lower bound.  Adding an
inverse-Bayesian \emph{relaxation} step (hypothesis renewal) to
ordinary Bayesian inference is by itself enough to render the dynamics
critical, with no external parameter adjustment.  We thus
show \BIB{} to be an inference algorithm exhibiting
internal-state criticality, operationally an extended on-off
intermittency on the simplex of hypothesis-space distributions.
Rather than self-organizing toward an absorbing critical state,
\BIB{} reaches criticality by continually reconstructing the
hypothesis-space boundary, a mechanism complementary to
self-organized criticality, one that makes the criticality robust across a
natural parameter range and opens a new direction for
the dynamics of inference models.
\end{abstract}

\maketitle

% =============================================================================

\section{Introduction}\label{sec:intro}

A convergence between criticality and inference dynamics has emerged
in recent years from both empirical and theoretical sides.
Empirically, neural recordings consistent with operation near a
critical point have been correlated with individual fluid
intelligence~\cite{ezaki2020}, cognitive task
performance~\cite{muller2025}, and genetic determinants of
cognition~\cite{xin2025}, within the broader picture of biological
systems poised near criticality~\cite{beggs2003,mora2011,munoz2018}.
Theoretically, maximum-entropy fits to multivariate neural data sit
near a critical surface of their parameter
manifold~\cite{tkacik2015,mastromatteo2011}, yielding Zipf-like
statistics without fine-tuning~\cite{schwab2014}, and predictive
perception admits a self-organized-instability
formulation~\cite{friston2012}.  These criticality signatures are often discussed in terms of
self-organized criticality
(SOC)~\cite{bak1987,bak1996,paczuski1996,dickman2000,pruessner2012,
watkins2016,markovic2014}, in which a balance between driving and
relaxation brings a system to a critical state without external
parameter tuning, producing scale-invariant statistics with universal
exponents.  What this picture nonetheless leaves open is the
\emph{mechanistic origin} of inference-side criticality.  The
empirical works cited above identify signatures of critical neural
dynamics and correlate them with cognition.  The theoretical works
argue, at the level of generic statistical-physical behavior, that
inferred models tend to lie near critical points.  Neither program
provides a concrete inference rule whose own dynamics is critical.
What is required is a minimal generative model in which criticality
emerges from the inference process itself, rather than being imposed
by external tuning.

At the behavioral level, the same convergence is sharpened by
\Levy{} walks, whose heavy-tailed displacement and dwell-time
distributions emerge near critical points of stochastic
dynamics and confer functional advantages for sparse-resource
search~\cite{viswanathan1999,abe2020}.  The same signature recurs
across the foraging and exploratory motion of biological agents such
as albatrosses~\cite{viswanathan1996,viswanathan1999}, marine
predators~\cite{sims2008,humphries2010}, and
humans~\cite{rhee2011,namboodiri2016}, placing the criticality of
inference dynamics at the intersection of mechanism and
behavioral strategy.  An explicit candidate for the inference
primitive identified above as missing is
\emph{Bayesian--inverse-Bayesian} (BIB) inference, introduced by
Gunji \textit{et al.}~\cite{gunji2017,gunji2018,gunji2021,gunji2026}:
the standard Bayesian update is augmented by a slower inverse step
that replaces the likelihood of the least-supported hypothesis with
the recent empirical pattern
($P(\delta\mid h)\!\leftarrow\!P(\delta)$)~\cite{gunji2017}.%
\footnote{Reference~\cite{gunji2017} introduces and formalizes this
inverse-Bayesian inference.  Refs.~\cite{gunji2018,gunji2021} develop
swarm-model realizations and Ref.~\cite{gunji2026} the most recent
formulation.}  Concrete
realizations include self-propelled swarms~\cite{gunji2021},
reward-based gameplay in the multi-strategy card game
Daihinmin~\cite{ibuka2024}, and related discrete settings, all of
which empirically recover heavy-tailed persistence statistics of
\Levy{} form.  A mathematical derivative of BIB was subsequently
developed by Shinohara
\textit{et al.}~\cite{shinohara2020,shinohara2020plos}, in which the
inverse step is recast as a slow forgetting--learning modification of
the highest-posterior hypothesis controlled by a forgetting rate
$\beta$ and a learning rate $\gamma$.  This derivative formulation
likewise produces power-law step-duration distributions in two-agent
continuous imitation simulations~\cite{shinohara2021,shinohara2022}.
Both formulations empirically generate heavy-tailed statistics, but
their relationship to SOC has been left explicitly open.  Shinohara
\textit{et al.}~\cite{shinohara2021} report that in their derivative
formulation ``the argument for SOC is only valid in the special case
where the learning rate is $\gamma=0.0$,'' and defer detailed SOC discussion to future work.
Whether either formulation corresponds to SOC in the
precise sense of statistical physics, and whether they share a
universality class if both are critical, thus remains a
community open question.  The present paper addresses this question
empirically for the original BIB formulation.  The parallel
characterization of the Shinohara derivative under the same protocol
is left for future work.

Practical artificial intelligence is concerned with quite different
things.  Where the statistical-physics view developed above asks
whether inference sits near a critical point, with its scale-free,
heavy-tailed (\Levy) signatures, game-playing AI seeks the opposite:
accuracy and convergence to an optimal strategy, as in the
Nash-targeting optimizers (counterfactual-regret minimization and its
variants~\cite{zinkevich2007,tammelin2014,brown2019,lanctot2009})
behind recent superhuman systems, for which criticality plays no role.
The present work sits at the intersection of the two.  Our testbed,
the symmetric two-player game of repeated rock-paper-scissors (RPS)
generalized to $N\in\{3,5,7,\dots\}$ hands in a cyclic-dominance
pattern, is the cleanest arena in which they meet.  Its Nash
equilibrium is trivial (uniform), so convergence-oriented optimizers
collapse to a random baseline, exposing the heavy-tailed structure
that a generative inference agent must instead capture.  RPS appears both as the canonical
substrate of spatial cyclic-dominance evolutionary
games~\cite{szolnoki2014}, where payoffs govern population-level
fitness and extinction, and as a canonical two-player adversarial
game in modern algorithmic game theory~\cite{neller2013}, in which
payoffs govern individual win/lose outcomes.  We adopt the cyclic
topology of the former in the adversarial setting of the latter.
The unique Nash mixed equilibrium of symmetric zero-sum $N$-hand RPS
is the uniform distribution $(1/N,\dots,1/N)$, on which those
Nash-targeting optimizers settle.  Recent large-language-model agents likewise apply
Nash-deviating heuristics rigidly with weaker adaptation to environmental change~\cite{zheng2025}.  Any
non-trivial dynamical structure observed in RPS-playing agents must
therefore originate outside game-theoretic optimization.  Human play,
however, departs systematically from this uniform-random baseline.
Direct observations of repeated human RPS report cycling patterns and
conditional responses to recent opponent
moves~\cite{wang2014,arai2025}.  Beyond RPS itself, similar departures
from rational-equilibrium baselines, such as win-streak persistence,
loss-induced strategy shifts, and other hot-hand/gambler's-fallacy
signatures, recur across iterated decision
tasks~\cite{dong2014,xuharvey2014,spanknebel2015}.  More broadly,
the inter-event times of human decisions across many substrates
follow heavy-tailed rather than Poisson
statistics~\cite{barabasi2005,vazquez2005,karsai2012,karsai2018},
attributed to the priority-driven nature of decision-making itself
rather than to environmental noise.  RPS therefore furnishes the simplest substrate in which
the Nash-optimal solution is trivial yet the empirical phenomenology
is non-trivial.  Bridging that gap calls for a generative inference
model rather than a Nash-targeting optimizer.

\begin{figure*}[t]
\centering
\includegraphics[width=\linewidth]{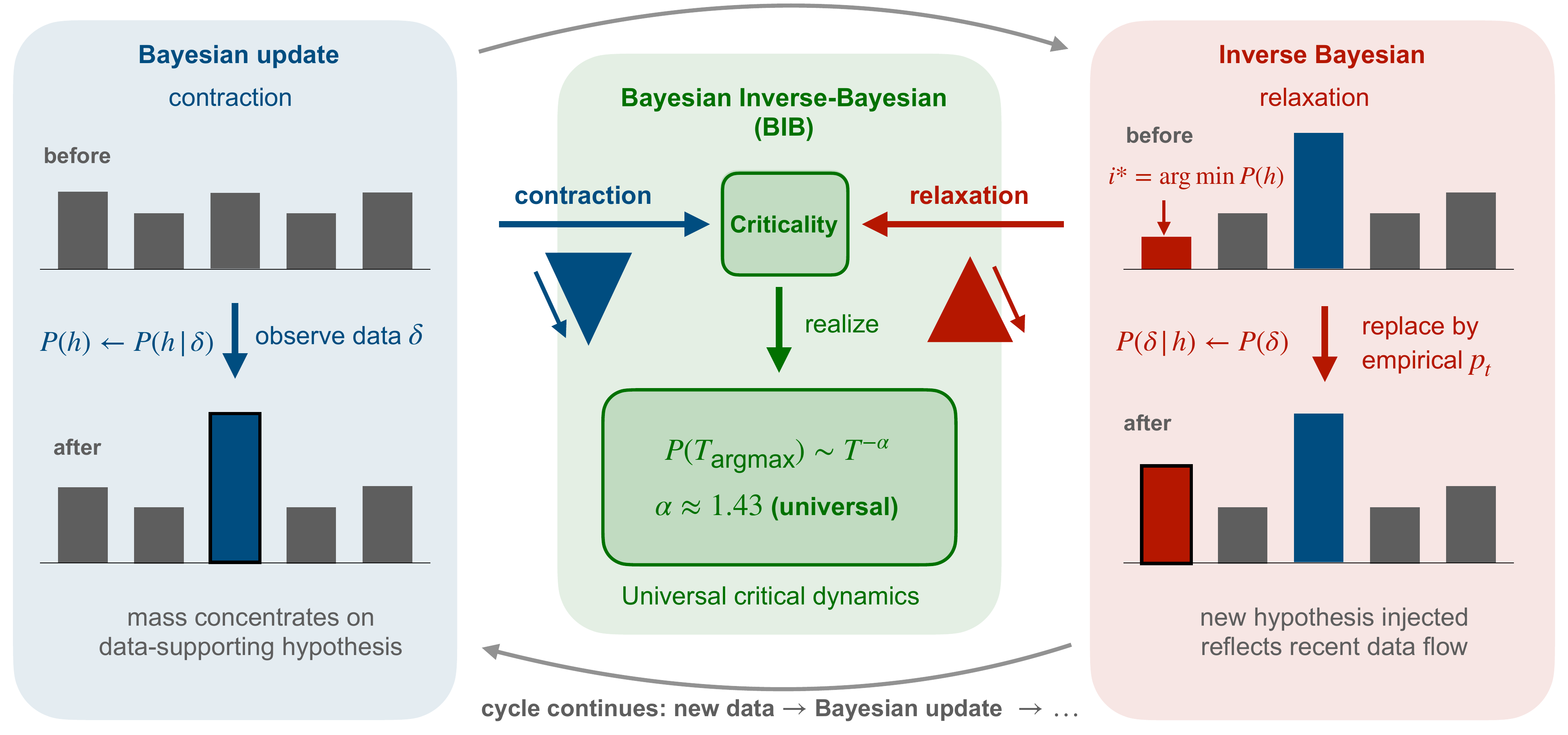}
\caption{Schematic of the Bayesian and inverse-Bayesian steps and the
resulting internal-state critical dynamics.  The Bayesian update (left)
contracts the posterior, concentrating mass on data-supporting
hypotheses.  The inverse-Bayesian replacement (right) renews the hypothesis set,
injecting a new hypothesis that reflects the recent empirical pattern.
Together, the Bayesian concentration and the inverse-Bayesian renewal
give rise to a critical state characterized by a
power-law persistence-time distribution
$P(\Targmax)\sim T^{-\alpha}$, where $\Targmax$ is the number of
consecutive steps the posterior's most-probable hypothesis remains
unchanged.}
\label{fig:soc-schematic}
\end{figure*}

We therefore devise a BIB inference model applicable to repeated
$N$-hand RPS gameplay, using the reward-based observation rule of
Ibuka and Sasai~\cite{ibuka2024} for direct comparability with the
most recent discrete-game implementation.  The model is dual in
nature.  It pursues an optimal strategy through the standard Bayesian
update of its hypothesis set, while its inverse-Bayesian step (which
replaces the lowest-posterior hypothesis with the empirical pattern of
recent observations) renews the hypothesis space and generates
heavy-tailed, scale-free dynamics inside its own representational
state (Fig.~\ref{fig:soc-schematic}).  This duality, strategy-seeking
on the surface and criticality-generating within, is what we propose
accounts for the human-versus-AI gap above.  The agent reproduces the
power-law signatures of human decision-making by virtue of its
internal mechanism, not by external fitting.  We demonstrate this in
$N$-hand RPS at large scale ($\sim\!10^{8}$ decision events in
total), establishing that \BIB{} reaches a robust internal critical state,
with no analogous universality under Bayes-only inference and with the
argmax-persistence distribution a power law of exponent
$\alpha\approx 1.43$, and reading the mechanism as
\emph{internal-state criticality} arising from continual
reconstruction of the hypothesis-space boundary.

Throughout, we keep the empirical findings, namely the measured
exponents and their robustness to design, hypothesis count, window
size, observation rule, and likelihood sharpness, separate from their
physical interpretation.  The measurements stand on their own, while
the internal-state-criticality reading is developed in
Sec.~\ref{sec:soc} and weighed against alternative explanations in the
discussion (Sec.~\ref{sec:behaviour}).  We stress at the outset that
the central claim is mechanistic rather than numerical.  The
inverse-Bayesian step pins the per-step log-posterior drift to zero as
a structural Kullback--Leibler fixed point of the update, and the
measured exponents are first-passage consequences of that driftless
walk.  The robust signatures are therefore the cross-design collapse
and the $\Nh$-invariance, not the convention-dependent value $1.43$.
Section~\ref{sec:model} introduces the BIB inference model on
$N$-hand RPS, including the reward-based update rule, agent types and
the experimental sweep.  Section~\ref{sec:univclass} reports the universality of the BIB
persistence exponent and its absence under Bayes-only updating, through
a direct comparison of the empirical distributions, followed by their
interpretation.  Section~\ref{sec:soc} develops the internal-state-criticality interpretation, including its on-off
intermittency identification, the finite-size scaling and
$\Nh$-rescaling analysis that tests it, observation-rule invariance,
and relation to the open SOC question.
Section~\ref{sec:robust} addresses robustness across observation
rule, game dimension, and window size.
Section~\ref{sec:reward} analyzes the reward statistics,
distinguishing internal-state universality from external
functionality.  Section~\ref{sec:behaviour} discusses behavioral
implications, limitations, and outlook.  Section~\ref{sec:conclusion}
concludes.

% =============================================================================
\section{Model: BIB inference on $N$-hand RPS}\label{sec:model}\label{sec:methods}

\subsection{$N$-hand cyclic-dominance RPS substrate}\label{sec:model-rps}

We consider the symmetric two-player zero-sum game of $N$-hand
rock-paper-scissors (RPS), generalized to a cyclic-dominance topology
in the standard way.  Let $\mathcal{D}=\{0,1,\dots,N-1\}$ denote
the set of $N$ hands, arranged on the cyclic order $\mathbb{Z}/N\mathbb{Z}$.  The win set
$\mathcal{W}\subseteq \mathcal{D}\times\mathcal{D}$ is
\begin{equation}
\mathcal{W} = \bigl\{(i,j) \;\big|\;
   j = (i+s)\bmod N,\; s=1,2,\dots,k\bigr\},
\end{equation}
with $k=(N-1)/2$ for odd $N$, which we adopt throughout to keep the
win/defeat relation symmetric.  Hand $i$ beats hand $j$ iff
$(i,j)\in\mathcal{W}$.  Standard $N=3$ RPS corresponds to $k=1$ and
the five-hand ``Rock, Paper, Scissors, Lizard, Spock''
extension~\cite{kass} to $N=5$, $k=2$.  We also test $N=7$, $k=3$.

We label the two players agent~A and agent~B.  For a single round,
the outcome from agent~A's perspective is
\begin{equation}
\mathrm{outcome}(d_A,d_B)=\begin{cases}
\mathrm{win}   & \text{if } (d_A,d_B)\in\mathcal{W},\\
\mathrm{defeat}& \text{if } (d_B,d_A)\in\mathcal{W},\\
\mathrm{tie}   & \text{otherwise,}
\end{cases}
\end{equation}
with reward $r_{\mathrm{win}}=+1$, $r_{\mathrm{defeat}}=-1$,
$r_{\mathrm{tie}}=0$.  We define win and defeat symmetrically through
the win set $\mathcal{W}$ and treat the tie as the residual case.  In
the present two-player game the residual reduces to $d_A=d_B$, and for
odd $N$ the cyclic-dominance relation is complete, so the two
non-residual cases are mutually exclusive and exhaustive over
$d_A\neq d_B$.  Casting the tie as the complement of the win/defeat
relations rather than as the single coincidence $d_A=d_B$ keeps the
outcome map well defined for the multi-player generalizations of the
substrate, where players may split across hands with no pairwise
decision.  At the uniform Nash equilibrium, both agents
realize expected reward $0$ per round.

\subsection{Reward-based BIB update rule}\label{sec:model-bib}\label{sec:methods-obs}

Following Ibuka and Sasai~\cite{ibuka2024}, we adopt a
\emph{reward-based} observation scheme in which each agent updates
its hypothesis posterior using an outcome-derived observation
$\delta\in\mathcal{D}$ rather than the opponent's hand directly:
\begin{equation}
\delta = \begin{cases}
d_A & \text{if outcome}=\mathrm{win},\\
\text{(no update)} & \text{if outcome}=\mathrm{tie},\\
\sim \mathrm{Uniform}\!\bigl(\mathcal{D}\setminus\{d_A\}\bigr) &
\text{if outcome}=\mathrm{defeat}.
\end{cases}
\label{eq:reward}
\end{equation}
On a win the agent reinforces its own successful hand by adopting it
as the observation.  On a loss it injects a counterfactual hand drawn
uniformly from the remaining $N-1$ hands
$\mathcal{D}\setminus\{d_A\}$.  On a tie it makes no update.  The rule
is therefore a reward-driven reinforcement-and-exploration heuristic
rather than a direct observation of $d_B$.  The agent never observes
the opponent's hand, and the win and loss branches differ in their
\emph{treatment} of the outcome (deterministic self-reinforcement
versus stochastic counterfactual exploration), not in the information
the round reveals about $d_B$.  That information is in fact symmetric
between win and loss.  Each non-tie outcome narrows $d_B$ to the
$k=(N-1)/2$ hands on the corresponding side of the cyclic-dominance
relation, coinciding only in the special case $N=3$ ($k=1$), where
either outcome fixes $d_B$ uniquely.

Each non-random agent maintains $\Nh$ hypotheses
$h_0,\dots,h_{\Nh-1}$, each associated with a likelihood vector
$L_i = P(\delta\mid h_i)\in\Delta^{N-1}$ and a posterior probability
$P(h_i)$.  The default $\Nh=10$ follows Ibuka and
Sasai~\cite{ibuka2024}.  We sweep $\Nh\in\{3,6,10,15,20\}$ in
Sec.~\ref{sec:results-nh} to test universality.

\paragraph{Hypothesis-space initialization.}  We compare two
initialization regimes.  In \emph{random initialization} each
likelihood vector $L_i$ is drawn from a half-normal distribution and
normalized as $L_{ij} = |Z_{ij}|/\sum_{j'}|Z_{ij'}|$ with
$Z_{ij}\sim\mathcal{N}(0,1)$.  In \emph{structured initialization} we
use, for $N=3$, ten template distributions corresponding to
interpretable hypotheses about the opponent: uniform
$(1/3,1/3,1/3)$, three concentrated $(0.8,0.1,0.1)$ permutations,
three biased $(0.5,0.25,0.25)$ permutations, and three
anti-concentrated $(0.10,0.45,0.45)$ permutations, mirroring the
structured initialization of Gunji \textit{et al.}~\cite{gunji2021}.
Both initializations begin with a uniform prior $P(h_i)=1/\Nh$.

\paragraph{Standard Bayesian update with conditional Jelinek--Mercer
regularization.}  Given an observation $\delta$ at time $t$, the
posterior is updated by Bayes' rule
\begin{equation}
P_{t+1}(h_i) = \frac{P(\delta\mid h_i)\,P_t(h_i)}{\sum_j P(\delta\mid h_j)\,P_t(h_j)} .
\end{equation}
To prevent posterior collapse, we apply a \emph{conditional}
smoothing step adapted from Ibuka and Sasai~\cite{ibuka2024}: whenever
$\min_i P(h_i)<0.002$, we set $P(h_i)\leftarrow 0.91\,P(h_i)+0.015$
and re-normalize.  This rule is mathematically equivalent to a
conditional Jelinek--Mercer
interpolation~\cite{jelinek1980,chen1996,manning1999} toward the
uniform reference $\pi_0(h_i)=1/\Nh$ with mixing weight
\begin{equation}
\alpha_{\mathrm{JM}}(\Nh)=\frac{0.015\,\Nh}{0.91+0.015\,\Nh},
\end{equation}
i.e.\ $\alpha_{\mathrm{JM}}\approx 0.05,\,0.09,\,0.14,\,0.20,\,0.25$
at $\Nh=3,\,6,\,10,\,15,\,20$ respectively.  The
\emph{conditional} triggering is essential to the dynamics:
Sec.~\ref{sec:results-nh} and Appendix~\ref{app:smoothing} show that
an unconditional always-on Jelinek--Mercer interpolation produces
qualitatively different statistics.

\paragraph{Inverse Bayesian step (BIB only).}  After history
accumulation reaches the window size $m$, the
Bayesian--inverse-Bayesian agent additionally performs a hypothesis
\emph{replacement}.  With
$\mathcal{H}_t=(\delta_{t-m+1},\dots,\delta_t)$ the window of the $m$
most recent \emph{emitted} observations, where ties emit no
observation and are skipped, as in Eq.~(\ref{eq:reward}), the empirical
histogram $p_t\in\Delta^{N-1}$ is
\begin{equation}
p_t(\delta) = \frac{1}{m}\sum_{\tau=t-m+1}^{t}
              \mathbf{1}_{\delta_\tau=\delta}.
\end{equation}
The likelihood vector of the lowest-posterior hypothesis is then
replaced by $p_t$~\cite{gunji2018}:
\begin{equation}
L_{i^{*}}\leftarrow p_t,\qquad i^{*}=\arg\min_i P(h_i).
\end{equation}
A Laplace-style additive smoothing~\cite{laplace1812,manning1999} is
applied conditionally to ensure positivity.  This is the Gunji-style
\emph{inverse Bayesian} operation.  The lowest-mass hypothesis is
replaced with the empirical pattern of recent observations,
\emph{creating new hypotheses} rather than merely re-weighting
existing ones.  We refer to this argmin-based update as the
\emph{argmin-replacement} mechanism of the original BIB formulation,
to distinguish it from the \emph{argmax-modification} mechanism of
the Shinohara derivative~\cite{shinohara2020,shinohara2020plos}, in
which the inverse step modifies the highest-posterior hypothesis
rather than replacing the lowest-posterior one.  The
two mechanistic labels are used in this sense throughout
Sec.~\ref{sec:soc}.

\paragraph{Action selection.}  At each step the agent selects its
hand by sampling (or selecting) a hypothesis and then sampling a hand
from its likelihood vector:
\begin{equation}
h^{*}\sim P(h),\qquad d\sim L_{h^{*}},\qquad \text{play } d.
\end{equation}
We compare two prediction modes: \emph{sample mode} (default),
$h^{*}\sim\mathrm{Categorical}(P(h))$, and \emph{argmax mode},
$h^{*}=\arg\max_i P(h_i)$.  The update is fully online.  The agent
carries no pre-trained parameters and undergoes no offline training
phase, so its dynamics arise entirely during play.

\subsection{Experimental sweep (six axes) and fitting protocol}\label{sec:model-sweep}\label{sec:methods-sweep}

Three agent types are compared: Random (uniform random hand,
no learning, mathematically equivalent to all CFR-family
algorithms~\cite{neller2013}), \BO (Bayes-only, Bayesian
update only, no inverse step), and \BIB
(Bayesian--inverse-Bayesian, adds the inverse step).  Three pair
conditions consist of two learner agents drawn from
$\{\BIB,\BO\}$ (\BIB-\BIB, \BO-\BO, \BIB-\BO).  The other three pair
either learner against random (\BIB-random, \BO-random) or random
against random, giving six pair conditions per design.

The six experimental axes are: game size $N\in\{3,5,7\}$, hypothesis
count $\Nh\in\{3,6,10,15,20\}$, window size $m\in\{10,20,50,100\}$,
design (\{random, structured\}$\times$\{sample, argmax\}), $\Pdh$
sharpness $\alpha_{\mathrm{init}}\in\{0.4,\dots,0.9\}$, and the six
opponent pairs $(a_A,a_B)$.
For the central design $\times$ window $\times$ pair sweep
($4\times 4\times 6=96$ conditions) the default values are $N=3$,
$\Nh=10$, $m=50$, $\alpha_{\mathrm{init}}=0.8$.  Other axes are
explored individually with the rest held at default.

For the argmax persistence distribution we fit three candidate
models using the \texttt{powerlaw}
package~\cite{alstott2014,clauset2009}: power-law (PL) with
$P(T)\propto T^{-\alpha}$, truncated power-law (TPL) with
$P(T)\propto T^{-\alpha}\exp(-\Lambda T)$, and exponential (Exp) with
$P(T)\propto \exp(-\Lambda T)$.  Model selection uses Akaike weights.
The ``best'' model is the one with maximum Akaike weight.  We adopt
the Clauset--Shalizi--Newman procedure~\cite{clauset2009} for
$x_{\min}$ selection (see also Refs.~\cite{stumpf2012,deluca2013}),
and report $\alpha_{\mathrm{TPL}}$ throughout.  Where the exponential is
preferred (as for some \BO{} laminar fits at large $\Nh$), we report
that fact rather than a power-law exponent.  The \Levy{} regime is
$1<\alpha\le 3$, with the foraging-optimal value at
$\alpha=2$~\cite{viswanathan1999}.  Full details of the simulation
scale, observables, and implementation are given in
Appendix~\ref{app:methods}.

% Methodological details (simulation scale, metrics, implementation)
% deferred to Appendix~B (app:methods).

% =============================================================================
\section{Universality of the persistence-time exponent}\label{sec:univclass}

\begin{figure*}[t]
\centering
\includegraphics[width=\linewidth]{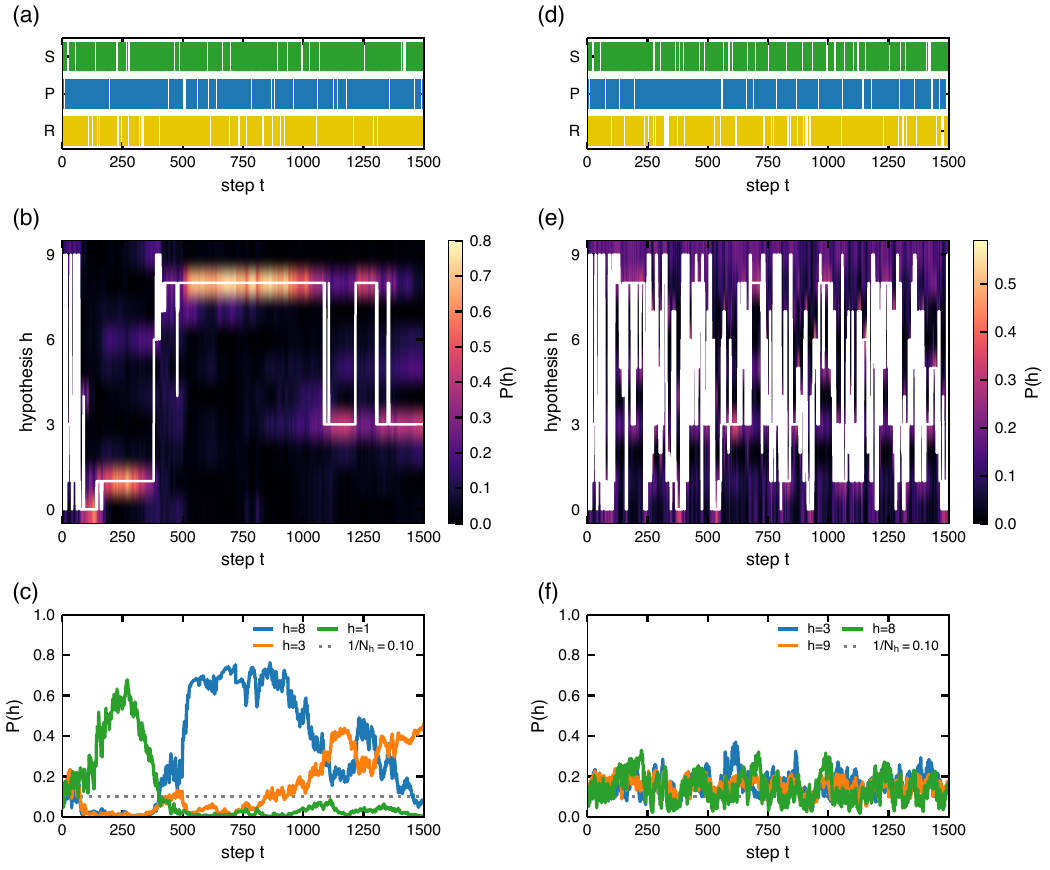}
\caption{Hypothesis-space dynamics, \BIB-\BIB{} (left column) versus
\BO-\BO{} (right column).  Same random seed, rs design (random init,
sample), $\Nh=10$, $m=50$, $T=1500$.  (a, d) Hand sequence of
agent~A (R yellow, P blue, S green).  (b, e) Posterior $P(h)$
heatmap with the argmax trajectory overlaid (white line).  (c, f)
Top-3 $P(h)$ trajectories with the uniform reference $1/\Nh$ shown
as a dotted line.}
\label{fig:dynamics}
\end{figure*}

Figure~\ref{fig:dynamics} contrasts the hypothesis-space dynamics of
the two inference schemes on a representative run.  \BIB{} (left column)
exhibits intermittent dominance of single hypotheses punctuated by
sudden reorganizations, whereas \BO{} (right column) stays close to a
mixed posterior throughout.  The remainder of this section quantifies
this qualitative contrast and shows that the heavy-tailed structure is
specific to the inverse-Bayesian dynamics.

\subsection{Universality of the BIB exponent}\label{sec:univclass-bib}\label{sec:results-univ}\label{sec:univclass-bo}\label{sec:results-bo}

Given the diversity of implementation choices across the BIB
literature (Sec.~\ref{sec:methods-sweep}), a natural starting
question is which of the four (initialization $\times$
prediction-mode) combinations should be regarded as canonical.  Our
results provide a meta-level answer.  Under \BIB, all four design
choices yield essentially the same persistence-time exponent, so the
choice is, within the range tested, empirically inconsequential.

\begin{figure}[tb]
\centering
\includegraphics[width=\linewidth]{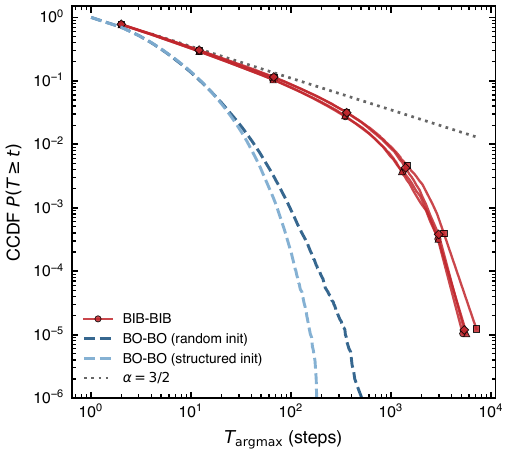}
\caption{Design-independent collapse of the \BIB{} argmax-persistence
exponent and its absence under Bayes-only updating.  Argmax-persistence
CCDFs, the four \BIB-\BIB{} designs (red) collapse onto a single heavy
tail ($\alpha\approx 1.43$), whereas \BO-\BO{} (blue, random- versus
structured-initialization families) departs from a power law at the
shortest persistence times.  $T=2\times 10^{5}\times\,20$ runs,
$\Nh=10$, $m=50$. The collapse, rather than the precise exponent value, is the primary signature of the universality class.
}
\label{fig:univ}
\end{figure}

Figure~\ref{fig:univ} shows the complementary cumulative
distribution functions (CCDFs) of the argmax persistence time
$\Targmax^{(A)}$ for the four \BIB-\BIB{} designs (rs/ra/ss/sa, color-
and marker-coded) and, for comparison, \BO-\BO{}, at the canonical
window size $m=50$.  The four \BIB-\BIB{}
curves lie almost on top of each other across nearly five
decades of probability density.  Truncated power-law fits yield
\begin{equation}
\alpha_{\BIB\text{-}\BIB}^{m=50}
= \{1.44,\;1.43,\;1.44,\;1.41\}
\label{eq:univ-arg}
\end{equation}
for designs (rs, ra, ss, sa) respectively, with cross-design range
$0.03$ (raw fit range $0.032$).  We adopt
$\alpha\approx 1.43$ as the canonical value of the
\emph{argmax-persistence} empirical universality class (a
condition-invariant robustness class, with the connection to the on-off-intermittency
universality class developed in Sec.~\ref{sec:soc}).  This cross-design
spread ($0.03$) is of the same order as the per-design fit
uncertainty, which a bootstrap ($B=200$ truncated-power-law refits per
design) places at interquartile ranges $\le 0.02$, so the four designs
are statistically \emph{consistent with a single exponent}, an
equivalence statement stronger than merely failing a test for
between-design difference.  This collapse is robust to the fitting
convention.  While the absolute exponent shifts with the $x_{\min}$
choice, the cross-design spread stays $\le 0.04$ under power-law,
truncated-power-law, and fixed-$x_{\min}$ variants alike
(Appendix~\ref{app:xmin}).

\paragraph{Argmax-persistent plateaux.}
The same collapse persists under a stricter \emph{plateau} observable
(maximal argmax-stable intervals with $P(h^{*})>\theta=0.3$ held for a
fraction $f=0.8$ of the run).  The \BIB-\BIB{} plateau-length CCDFs again
collapse onto a single heavy-tailed family, whereas \BO-\BO{} does not
(Appendix~\ref{app:plateau}).

The inverse-Bayesian step is necessary for this behavior.  The Bayes-only
control, which applies the same Bayesian update without the renewal step, does
not form a comparable power law.\footnote{Shinohara \textit{et al.}~\cite{shinohara2021} observed a ``universality near $\eta\approx 2$'' in two-agent BIB-style continuous imitation.  In the present discrete reward-based setting that value is not recovered as a stable exponent, indicating that it reflects the specific continuous-imitation regime studied there rather than a universal signature of the framework.}  Under the identical model-selection procedure
(Clauset $x_{\min}$ with Akaike weights over power-law, truncated-power-law,
exponential, lognormal, and stretched-exponential models,
Appendix~\ref{app:modelselect}), the argmax-persistence distribution of \BO-\BO{} does not satisfy the
heavy-tail criterion at any design.  For random initialization the preferred description is a stretched
exponential or lognormal.  For structured initialization the best-fitting
truncated power law spans less than $1.2$ decades, in both cases failing the
requirement of a heavy-tailed fit over at least two decades.  The same holds
across the hypothesis-count sweep for all $\Nh\ge 6$.  The limited
design-dependence that \BO{} does show is confined to the initialization.
Lacking the renewal that would otherwise overwrite them, the Bayes-only agent
retains the information content of its initial likelihoods, so the random- and
structured-initialization families remain distinct (Fig.~\ref{fig:univ}),
while the prediction mode has little effect.

\subsection{Specificity among adaptive learners}\label{sec:learners}

The \BIB-versus-\BO{} contrast shows that the inverse step, not
Bayesian updating, is what creates the critical class.  A complementary
question is whether that criticality is specific to \BIB{} or generic
to adaptive play on these uniform-Nash games.  We therefore compared
\BIB{} with three standard, parameter-light learners under the
identical fit pipeline (Appendix~\ref{app:xmin}): win-stay/lose-shift
(WSLS)~\cite{nowak1993}, tabular $Q$-learning~\cite{watkins1992}, and
regret matching~\cite{hartmascolell2000}, each run against a
uniform-random opponent and in self-play.  For every agent we read the
persistence of its dominant internal preference (the argmax hypothesis
for \BIB{}, the argmax cumulative regret for regret matching) or, for
the reactive baselines that carry no internal preference, the
played-action run length.

\begin{figure*}[tp]
\centering
\includegraphics[width=\linewidth]{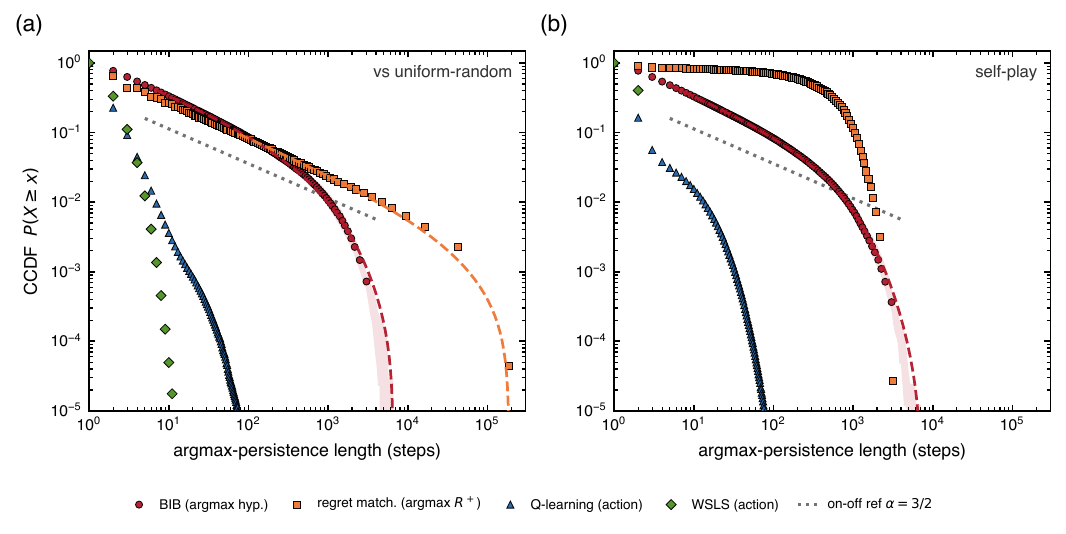}
\caption{Specificity of the critical class to the inverse-Bayesian
rule among adaptive learners.  Internal argmax-persistence CCDFs of
\BIB{} (red circles, shaded band over the four designs rs/ra/ss/sa)
versus standard learners at $\Nh=10$, $m=50$, $T=2\times10^{5}$
($40$ seeds), showing the played-action run length for the reactive
baselines $Q$-learning (blue triangles) and WSLS (green diamonds) and
the argmax cumulative regret for regret matching (orange squares).
(a)~Against a uniform-random opponent.  (b)~In self-play.  Dashed
curves are truncated-power-law fits.  The dotted guide is the on-off
$\alpha=3/2$ slope.}
\label{fig:learners}
\end{figure*}

Against a uniform-random opponent [Fig.~\ref{fig:learners}(a)], regret
matching is critical in the same $3/2$ class as \BIB{}.  Its cumulative
regrets are driftless random walks, so the persistence of the
maximum-regret action is a truncated power law with interior exponent
$\alpha\approx 1.50$, the Sparre--Andersen first-return value, while
\BIB{} gives $\alpha\approx 1.45$.  WSLS dwell times are geometric and
$Q$-learning's are lognormal.  Neither shows a scale-free regime.  This
regret-matching coincidence is, however, imposed by the neutral
environment rather than self-organized, and it vanishes under self-play
[Fig.~\ref{fig:learners}(b)]: once the opponent adapts, the regret walk
acquires a drift and its tail loses the scale-free form, whereas
\BIB{} retains $\alpha\approx 1.43$.  Only \BIB{} stays critical when
the opponent is itself adaptive.

The distinction does not rest on the bare exponent, which is the
generic first-passage value shared by any driftless walk, but on
robustness.  Only \BIB{} is critical across both opponent class and
hypothesis count $\Nh$ (Sec.~\ref{sec:fss-collapse}).  Regret matching
is critical only against a neutral opponent.  A credible $3/2$ claim
further demands a power-law range of several decades with a stable
exponent, namely a small gap between the pure- and truncated-power-law
fits, which \BIB{} satisfies ($\approx 3$ decades, gap
$\lesssim 0.2$) and the reactive baselines do not ($\approx 1$ decade,
or a strongly curved tail).  The internal critical state is therefore
produced by the inverse-Bayesian step, not by adaptive play as such.
Full statistics are in Appendix~\ref{app:rlbaselines}.

The contrast is itself nontrivial.  Standard Bayesian inference is a
smooth, parameter-sensitive process, so one would expect the inverse
step merely to shift the dynamics rather than to produce a
qualitatively different statistical regime.  What we observe instead
is a sharp separation: \BIB{} realizes a \emph{single} universality
class decoupled from initialization and $\Nh$, whereas Bayes-only
updating produces no comparable power law at all.  The mechanism behind
this separation, and its place within established statistical-physics
frameworks, is the subject of the following section.

% -----------------------------------------------------------------------------

% =============================================================================
\section{Internal-state criticality from hypothesis renewal}\label{sec:soc}\label{sec:disc-soc}

In this section we analyze the universal critical mechanism of \BIB{}.  We characterize this internal-state criticality at three levels: as a critical state of a single driftless log-posterior walk, identified operationally with on-off intermittency, and reached not through an absorbing-state instability but through continual reconstruction of the hypothesis-space boundary.

\subsection{Hypothesis renewal and the first-return prediction}\label{sec:soc-mech}

BIB combines two operations with opposing tendencies.  The Bayesian
update is a contraction that concentrates posterior mass on
data-supporting hypotheses and would lead to posterior collapse if
applied alone.  The inverse-Bayesian step is a relaxation that replaces
the lowest-posterior hypothesis with the empirical histogram of the
recent $m$ observations~\cite{gunji2017}, and would prevent any
sustained accumulation if applied alone.  A two-hypothesis reduction
makes the resulting dynamics explicit.  Writing $p=P(h_1)$ and the
log-odds $u=\log[p/(1-p)]$, Bayes' rule
$p\mapsto L_1 p/[L_1 p + L_2(1-p)]$ is purely additive,
\begin{equation}
u_{t+1}=u_t+\eta_t,\qquad
\eta_t=\log\frac{L_1(\delta_t)}{L_2(\delta_t)},
\label{eq:logodds-walk}
\end{equation}
with $\eta_t$ the per-step log-likelihood ratio, and the normalizing
evidence is common to both hypotheses and cancels.  At the symmetric
fixed point no hypothesis is favored, so $\langle\eta_t\rangle=0$ and
$u_t$ is an unbiased random walk.  The argmax changes when $u$ first
returns to zero, so the argmax-persistence time is a first-return time, the canonical
case of a first passage of the walk to a threshold.
The laminar phase, defined by single-hypothesis dominance and analyzed
in Sec.~\ref{sec:soc-onoff}, is the first passage of the same walk to a
different threshold.  The
argmax persistence of Sec.~\ref{sec:univclass} and the laminar phase
are therefore dual readings of one first-passage process, which is why
a single class organizes both, and
for an unbiased walk with symmetric increments the Sparre--Andersen
theorem fixes the tail universally at $P(T)\sim T^{-3/2}$.

The vanishing drift is not merely a consequence of the two-state
symmetry, but a fixed point that the inverse step enforces.  Taking
the expectation of the increment over the data-generating distribution
$p$ gives the identity
\begin{equation}
\langle\eta\rangle = D_{\mathrm{KL}}(p\,\|\,L_j) - D_{\mathrm{KL}}(p\,\|\,L_i),
\label{eq:kldrift}
\end{equation}
with $L_i,L_j$ the likelihoods of the dominant and competing
hypotheses, the leader gaining log-posterior at a rate set by how much
KL-closer its likelihood lies to the data.  With fixed likelihoods the
dominant hypothesis is KL-closest, so $\langle\eta\rangle>0$ and the
posterior concentrates without bound.  The inverse-Bayesian step breaks
this by re-injecting the least-supported hypothesis as the recent
empirical histogram, whose expected divergence from the data is the
finite-sample value
$\mathbb{E}[D_{\mathrm{KL}}(p\,\|\,\hat p)]\approx (N-1)/(2m)$~\cite{coverthomas2006}.
A competitor at this near-zero divergence is continually available, so
whenever the leader opens a gap a freshly renewed hypothesis competes
it down, pinning the stationary drift $d\equiv\langle\eta\rangle$ at zero as
a fixed point of the inference rather than an imposed condition.  In the
production model the within-phase drift falls from
$d=+0.0044\pm 0.0008$ at renewal probability $q=0$ to
$|d|<10^{-4}$ for every $q\ge 0.05$, so the zero-drift
state is set by the presence of the renewal, not its rate
(Appendix~\ref{app:heuristic}).  The reduction predicts the following
phenomenology: the $3/2$ first-return law on the $d=0$ line, a
systematic offset of the measured exponent set only by the
finite-sample residual $(N-1)/(2m)$ and vanishing as $m\to\infty$, and
invariance to the hypothesis count $\Nh$, whose posterior-spread
scaling the renewal absorbs (Sec.~\ref{sec:fss-nhinv}).  A control
comparison confirms the role of the directed renewal.  Replacing it with
sequential importance resampling or a prior-restart does not reproduce
the on-off laminar structure (Appendix~\ref{app:resampling}).

\subsection{Empirical confirmation of the on-off exponent}\label{sec:soc-onoff}

\begin{figure}[tbp]
\centering
\includegraphics[width=\linewidth]{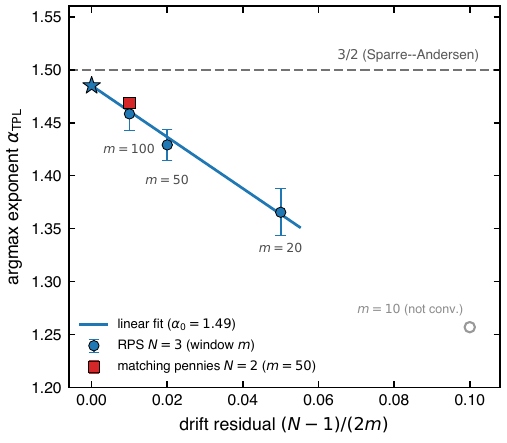}
\caption{Extrapolation of the argmax exponent to $3/2$ as the
finite-sample drift residual vanishes.  Measured \BIB-\BIB{}
argmax-persistence exponent $\alpha_{\mathrm{TPL}}$ (truncated-power-law
fit over the $[5,10^{4}]$ power-law region, both agents pooled) versus
the renewal residual $(N-1)/(2m)$.  Blue circles, rock--paper--scissors
($N=3$) at observation windows $m\in\{20,50,100\}$ (error bars, SD over
the four designs).  Red square, matching pennies ($N=2$, $m=50$), which
halves the residual at fixed $m$.  The two routes coincide at equal
residual, and the line is a least-squares fit through the converged points,
extrapolating to $\alpha_0\approx 1.49$ at zero residual (star), close to
the Sparre--Andersen $3/2$ (dashed).  The $m=10$ window (open symbol) is
shown for reference but excluded from the fit, its tail being too short
for a converged slope.}
\label{fig:drift-residual}
\end{figure}

The reduction of Sec.~\ref{sec:soc-mech} places the BIB persistence
statistics in the on-off intermittency class of Platt, Spiegel and
Tresser~\cite{platt1993,heagy1994}, in which a dynamical system whose
invariant manifold is destabilized by a stochastically modulated
parameter alternates between long laminar phases near the manifold and
brief bursts away from it.  The asymptotic distribution of laminar
phase lengths is universal,
\begin{equation}
P(\tau_{\mathrm{lam}}) \;\sim\; \tau_{\mathrm{lam}}^{-3/2}
\qquad\text{as }\tau_{\mathrm{lam}}\to\infty,
\label{eq:onoff-universal}
\end{equation}
the $-3/2$ exponent being robust to driving-noise
details~\cite{heagy1994,pikovsky2001} and verified
experimentally~\cite{hammer1994}.  In our setting the invariant
manifold is the high-$\max_h P(h)$ regime in which a single hypothesis
dominates the posterior, and the stochastic driver is the opponent-hand
sequence together with the empirical histogram injected by the
inverse-Bayesian step.  We use this on-off correspondence operationally,
through the shared $3/2$ first-return law and the laminar/burst
phenomenology, rather than through an explicit construction of the
invariant subspace and a transverse Lyapunov exponent crossing zero at
a blowout bifurcation, which we leave to future work
(Sec.~\ref{sec:limitations}, limitation~(iv)).  At the working $\Nh=10$ the same
first-return mechanism operates among the $\Nh-1$ competitors.  We measure the laminar phase
directly, the continuous interval during which $\max_h P(h)>\theta$ at
threshold $\theta=0.4$, well above the uniform value $1/\Nh$.  The
empirical numbers support the identification quantitatively.  The
\BIB-\BIB{} laminar exponent
\begin{equation}
\begin{split}
&\alpha^{\mathrm{lam}}_{\BIB\text{-}\BIB} = 1.325 \pm 0.006 \\
&\text{(cross-design SD over the $4$ designs at $\theta=0.4$)}
\end{split}
\label{eq:bib-lam}
\end{equation}
lies within $12\%$ of the universal value $\alpha = 3/2$ of
Eq.~(\ref{eq:onoff-universal}), with a cross-design range of $0.014$.  This exponent varies by no more than $0.06$ across the
full $P(d\mid h)$ sharpness sweep ($2$ structured designs ss, sa
$\times\,6$ sharpness values, Sec.~\ref{sec:robust-sharp}), comparable
to the residual scatter of the
universal exponent in classical on-off intermittency
experiments~\cite{hammer1994}.  The Bayes-only control does not develop
this on-off laminar structure.  Under the same model-selection procedure
(Appendix~\ref{app:modelselect}) the \BO-\BO{} laminar-phase
distribution fails the heavy-tail criterion at every design, and the
contrast is direct in the dynamics, where at $\theta=0.4$ the \BO-\BO{}
laminar phases occupy about one quarter of the run with a mean length
of four to five steps, against about one half and forty-two steps for
\BIB-\BIB{}, the Bayes-only posterior maximum rarely exceeding the
threshold.  The Bayes-only agent stays close to a mixed posterior and
does not sustain the heavy-tailed renewal, so the inverse step is
necessary for the on-off mechanism.  The
argmax-persistence exponent
$\alpha^{\mathrm{arg}}_{\BIB\text{-}\BIB} = 1.43 \pm 0.02$ sits in the
same regime.  The multi-hypothesis competition and the finite-window
correlations account for the slight excess of the pure power-law fit to
argmax persistence ($\alpha_{\mathrm{PL}}\approx 1.59$) above $3/2$.  This
fit overestimates the slope by neglecting the cutoff, whereas the
truncated-power-law fit reported throughout sits just below $3/2$.  This
systematic shift from $3/2$ has a single origin, the finite observation
window $m$.  The inverse-Bayesian renewal re-injects the renewed
hypothesis near the data at an expected divergence that scales as
$(N-1)/(2m)$ (Appendix~\ref{app:heuristic}).  This residual does not
reintroduce drift, which the renewal pins to zero throughout
($|d|<10^{-4}$).  It acts instead by correlating successive increments
of the otherwise independent log-posterior walk and placing the
dynamics in the additive--multiplicative on-off regime, shifting the
effective exponent away from the independent-increment
Sparre--Andersen value $3/2$ and vanishing only as $m\to\infty$.  Two independent lines of evidence fix this as the
cause.  First, the exponent rises monotonically toward $3/2$ as the
window grows, reaching a mean of $1.458$ at $m=100$
(Sec.~\ref{sec:robust-mN}).  Second, reducing the game size to the
two-action matching-pennies game ($N=2$) halves the residual at fixed
$m$ and shifts the exponent measurably closer to $3/2$
(Sec.~\ref{sec:robust-mp}).  Plotted against the residual, the window and
alphabet routes collapse onto one trend extrapolating to $3/2$
(Fig.~\ref{fig:drift-residual}).  Notably, the cause is the window $m$ and
the alphabet size $N$, \emph{not} the hypothesis count $\Nh$, consistent
with the $\Nh$-invariance established in Sec.~\ref{sec:fss-nhinv}.  This
finite-$m$ correction is \emph{realized through}, rather than
competing with, the additive--multiplicative on-off
regime~\cite{nakao1998,venkataramani1996} (the extended on-off
intermittency discussed by Bertin~\cite{bertin2011}): because the
inverse-Bayesian replacement combines multiplicative reweighting with
additive empirical-histogram injection, the mixture places the dynamics
in precisely the regime in which the on-off exponent departs from
$3/2$, so the measured value is the finite-$m$ effective exponent of
that mixed class rather than a separate constant.  Finite $m$ is the
cause, and the additive--multiplicative class is the form the correction
takes, not an independent explanation.

% -----------------------------------------------------------------------------

\subsection{$\Nh$-invariance and the absorption mechanism}\label{sec:fss}\label{sec:fss-nhinv}\label{sec:results-nh}\label{sec:fss-spread}\label{sec:results-extra}\label{sec:soc-nhinv}

\begin{figure*}[tp]
\centering
\includegraphics[width=\linewidth]{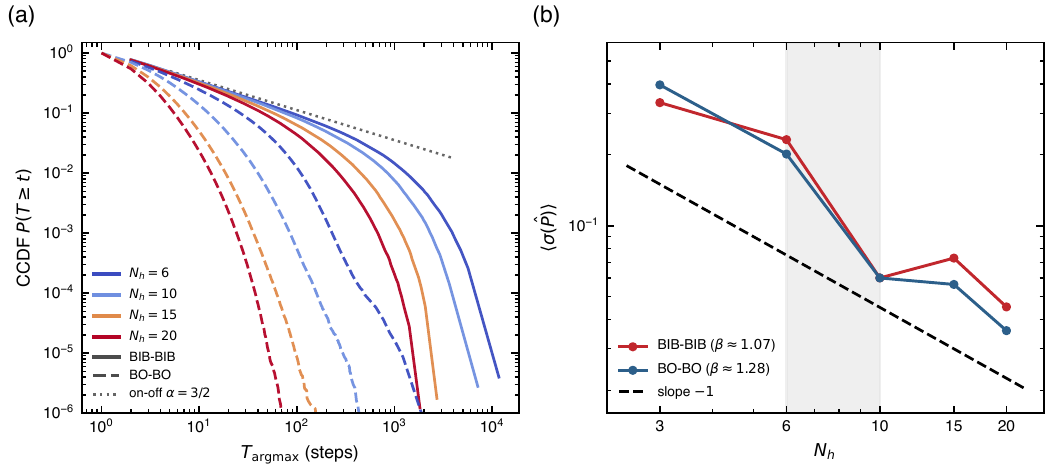}
\caption{The $\Nh$ sweep.  (a)~Argmax-persistence CCDFs for \BIB-\BIB{} and
\BO-\BO{} at $\Nh\in\{6,10,15,20\}$ (eight curves, each pooled over
designs, agents, and runs).  The \BIB{} curves overlap across the core
regime $\Nh\in\{6,10\}$, confirming $\Nh$-invariance of the exponent
($\alpha\approx 1.43$), and depart only at the finite-$\Nh$ boundary
$\Nh\ge 15$.  The \BO{} curves are not power laws at any $\Nh$ and shift
to shorter timescales as $\Nh$ grows.  Gray dotted, on-off-intermittency
reference $\alpha=3/2$.  (b)~Time-averaged posterior spread $\sigma(\Ph)$
versus $\Nh$ (log--log), pooled over designs, for \BIB-\BIB{} and
\BO-\BO{} (two curves) with power-law fits and the idealized slope $-1$,
with the core regime $\Nh\in\{6,10\}$ shaded.  \BIB{} has
spread-scaling exponent $\beta\approx 1.07$ and \BO{} a steeper
$\beta\approx 1.28$, with no comparable \BO{} power law.}
\label{fig:nhswp}
\end{figure*}

The dependence of the BIB exponent on the number of hypotheses $\Nh$
tests whether the exponent is tuning-free or tied to a particular
$\Nh$.  We ran large-scale
simulations at $m=50$, $N=3$ for $\Nh\in\{3,6,10,15,20\}$ across all
four designs.  Figure~\ref{fig:nhswp}(a) overlays the persistence CCDFs across $\Nh$
for \BIB-\BIB{} and \BO-\BO{}.

The four \BIB-\BIB{} curves coincide tightly across all four designs
in the \emph{core central regime} $\Nh\in\{6,10\}$ (shaded band):
the per-design two-point means are $\{1.438,\,1.426,\,1.450,\,1.418\}$
for (rs, ra, ss, sa), with cross-design pooled
mean $\alpha=1.433$ and sample standard deviation $0.014$ across
$n=8$ measurements.

We refer to $\Nh\in\{6,10\}$ as the \emph{core central regime} of the
BIB universality and adopt $\alpha=1.43$ as the canonical
value.  Outside this regime, finite-$\Nh$ effects appear and become
design-specific.  At $\Nh=3$, the degenerate limit, all four designs converge to
$\alpha\approx 1.62$.  At $\Nh\in\{15,20\}$ the persistence distributions
no longer satisfy the heavy-tail criterion (spanning under two decades,
Appendix~\ref{app:modelselect}), so no exponent is quoted there.  The
curves stay close to the core behavior at $\Nh=15$ and begin to depart at
$\Nh=20$, with the departure concentrated in the rs design.  This loss of the
heavy-tail verdict at large $\Nh$ is a dynamic-range effect rather than
a change of class.  Since the cutoff scales as $T_{\max}\propto\Nh^{-z}$
(Sec.~\ref{sec:fss-collapse}), the power-law window falls below two
decades as $\Nh$ grows, so the exponent becomes unmeasurable before it
would become non-universal, consistent with the curves staying parallel
to the core slope wherever they can still be read.  The \BO-\BO{} curves do not exhibit a regime of
universality at any $\Nh$.  Under the same model selection the \BO-\BO{} distributions fail the
heavy-tail criterion at every design for all $\Nh\ge 6$ (best fits
stretched-exponential, lognormal, or truncated power laws spanning under
two decades), so \BO{} has no power-law exponent to track across $\Nh$.
Only at the degenerate $\Nh=|\mathcal{D}|=3$ do some designs pass.

To
verify that the $1/\Nh$ scaling is operating in our system, we
directly measure the per-step posterior spread
$\sigma(\Ph)=\mathrm{std}_i P(h_i)$ for each agent across all four
designs.  Panel~(b) of Fig.~\ref{fig:nhswp} shows the
time-averaged $\sigma$ as a function of $\Nh$ for \BIB-\BIB{}
and \BO-\BO{}, pooled over the four designs.  Both pairs exhibit the
predicted $\sigma\propto \Nh^{-\beta}$ scaling.  The \BIB-\BIB{}
per-design exponents are $\beta=\{1.06,\,1.07,\,1.06,\,1.08\}$ with
cross-design mean $\beta_{\BIB}=1.067\pm 0.008$.  The \BO-\BO{}
per-design exponents are $\beta=\{1.28,\,1.26,\,1.31,\,1.27\}$ with
mean $\beta_{\BO}=1.278\pm 0.020$.

Importantly, the same $\sigma\propto\Nh^{-\beta}$ scaling is present in
both \BIB{} and \BO{}, in all four designs.  \BIB{} and Bayes-only
updating differ only in whether this scaling is absorbed into the
hypothesis-renewal dynamics.  In \BO{}, no inverse step exists, so the
persistence statistics track how often the dominant hypothesis flips:
smaller $\sigma$ means closer ties and more frequent flipping, and the
distribution accordingly fails to settle into a power law, shifting to
shorter timescales as $\Nh$ grows (Appendix~\ref{app:modelselect}).  In
\BIB{}, by contrast, the inverse step provides a relaxation channel
that absorbs the $\sigma$ scaling into the renewal dynamics, so that the
persistence-time distribution retains its shape and only the upper
cutoff slowly shrinks, by a factor of $\sim 16$ over the same $\Nh$
range, leaving the exponent invariant.  $\Nh$ thus acts as an inverse
finite-size variable.  Unlike a spatial system size, where the cutoff
grows with the number of degrees of freedom, a larger hypothesis set
here yields a finer posterior ($\sigma\propto\Nh^{-\beta}$) and a
shorter cutoff, the renewal absorbing the $\sigma$-scaling so that the
invariant is the exponent rather than the cutoff.  Consistently, \BIB{}'s
$\beta\approx 1.07$ sits very close to the information-theoretic lower
bound $\beta\geq 1$ for an equilibrated $\Nh$-element distribution,
whereas \BO{}'s $\beta\approx 1.28$ exceeds it by $0.28$, the excess
decay attributable to ongoing posterior concentration in the absence
of any renewal mechanism.  The difference between \BIB{} and Bayes-only
updating thus lies not in the underlying scaling of the posterior but
in whether the inverse step absorbs it.

Structurally, the argmin-replacement BIB rule shares the ingredients
commonly associated with
SOC~\cite{bak1987,bak1996,dickman2000,pruessner2012,watkins2016,munoz2018,markovic2014}:
two opposing operations, criticality reached without external tuning,
and parameter-independent power-law statistics.  The criticality lives
entirely in the simplex of probability distributions over hypotheses,
an internal representational space rather than a spatial array.  How
this internal-state criticality relates to, and departs from, classical
SOC is taken up in Sec.~\ref{sec:soc-shinohara}.  We argue that it
arises by continual reconstruction of the hypothesis-space boundary
rather than by self-organization toward an absorbing state.

\subsection{Finite-size scaling collapse and cutoff exponents}\label{sec:fss-collapse}\label{sec:results-fss}

A tuning-free critical structure should obey finite-size scaling, under
which the persistence-time distribution collapses onto a single curve
once the control parameter and the cutoff are
rescaled~\cite{privman1984,pruessner2012}.  The $\Nh$ sweep
(Sec.~\ref{sec:results-nh}) tests this directly, with $\Nh$ as the
finite-size control.

\paragraph{Scaling ansatz.}  At a critical point, the
persistence-time distribution should satisfy
\begin{equation}
P(\Targmax\mid \Nh) \;=\; T^{-\alpha}\,
\mathcal{F}\!\bigl(T\,\Nh^{\,z}\bigr),
\qquad T\equiv \Targmax,
\label{eq:fss}
\end{equation}
where $\alpha$ is the (universal) critical exponent, $z$ is a
\emph{finite-size scaling exponent} that sets the dependence of the
cutoff $T_{\max}(\Nh)$ on the control parameter and $\mathcal{F}$ is
a universal scaling function with $\mathcal{F}(x)\to\mathrm{const}$
for $x\to 0$ and $\mathcal{F}(x)\to 0$ exponentially for
$x\to\infty$.

\paragraph{Cutoff scaling.}  The cutoff time follows a single power of
$\Nh$ across the sweep.  Fitting the truncated power-law cutoff $1/\Lambda$ at the
representative rs design for $\Nh\in\{3,6,10,15,20\}$ gives the
sequence $1/\Lambda = \{6879,\,2751,\,1373,\,591,\,435\}$, well
approximated by $1/\Lambda \propto \Nh^{-z}$ with
\begin{equation}
z_{\BIB} \;=\; 1.49\pm 0.07 \qquad (R^{2} = 0.99).
\label{eq:zfss}
\end{equation}
The corresponding \BO-\BO{} cutoff is not described by a single power of
$\Nh$ and is better fit by an exponential at the largest $\Nh$,
consistent with \BO{} lacking a finite-size-scaling description.

\paragraph{Scaling collapse and the critical window.}  The implicit
collapse of the \BIB{} CCDFs is visible in Fig.~\ref{fig:nhswp}(a),
where across the core central regime $\Nh\in\{6,10\}$ the curves are
parallel.  An explicit rescaling $T\to T\,\Nh^{z_{\BIB}}$ overlaps them
onto a single curve within line thickness across nearly four decades of
probability density, the cross-$\Nh$ deviation remaining below $5\%$ in
the bulk.  The collapse degrades at the two boundaries.  At the
degenerate limit $\Nh=|\mathcal{D}|=3$ all four designs shift together
to $\alpha\approx 1.62$, a design-independent edge.  At the largest
$\Nh$ the fits no longer satisfy the heavy-tail criterion, with the
departure concentrated in the rs design while the others stay close to
the core behavior.  Since the posterior-spread exponent $\beta$ remains
common to all four designs there, we read this as a finite-$\Nh$
boundary of the universal regime rather than a crossover to \BO.  The
universal regime thus occupies a finite window of $\Nh$ above the
observation-alphabet size $|\mathcal{D}|$, consistent with the
game-dimension sweep of Sec.~\ref{sec:robust-mN}, which holds $\Nh=10$
above $|\mathcal{D}|$ for every $N\in\{3,5,7\}$.

\paragraph{Observable-dependent finite-size cutoffs.}  The cutoff
exponent of Eq.~(\ref{eq:zfss}) characterizes the argmax-persistence
observable.  Run on that observable the $1/\Lambda$-cutoff pipeline
reproduces $z^{\mathrm{arg}}\approx 1.49$, consistent with
Eq.~(\ref{eq:zfss}).  Applying the \emph{identical} pipeline to the
laminar-phase length across $\Nh\in\{3,6,10,15,20\}$ yields a markedly
steeper laminar cutoff scaling,
\begin{equation}
z^{\mathrm{lam}}_{\BIB} = 2.44 \pm 0.10 \qquad (R^{2} = 0.995),
\label{eq:zlam}
\end{equation}
(or $2.34\pm 0.09$ excluding the $\Nh$-boundary points), separated from
the argmax value by more than six standard deviations even though the
laminar tail itself stays in the $3/2$ class (core
$\alpha^{\mathrm{lam}} = 1.325\pm 0.01$, the SD here taken over the
$\Nh\in\{6,10\}$ core rather than over designs at fixed $\Nh$ as in
Eq.~(\ref{eq:bib-lam})).  We read
this as the two observables recording first-passage times of the
\emph{same} driftless log-posterior walk to \emph{different} absorbing
conditions, namely a change of the argmax incumbent versus the loss of
single-hypothesis dominance ($\max_h P(h) < \theta$), so that the
shared $3/2$ tail coexists with observable-dependent finite-size
cutoffs.  This is the first-passage picture of
Sec.~\ref{sec:soc-mech}, now with $\Nh$ setting the size of the basin
from which the walk must escape, while the escape condition, not the bulk
dynamics, fixes the cutoff exponent.  The laminar measurement is a
method-matched re-analysis.  A full-scale rerun would refine its value,
but the $z^{\mathrm{lam}} \gg z^{\mathrm{arg}}$ separation is robust.
The cutoff exponent is therefore a non-universal, boundary-dependent
quantity, whereas the universality resides in the shared $3/2$ tail.

% =============================================================================

% =============================================================================
\subsection{Relation to the open SOC question}
\label{sec:soc-shinohara}

The relationship between BIB-type criticality and self-organized
criticality was left explicitly open by Shinohara
\textit{et al.}~\cite{shinohara2021} (Sec.~\ref{sec:intro}).  We read the
BIB mechanism as complementary to absorbing-state
self-organized criticality rather than as an instance of it.  In its
canonical (sandpile) form, SOC is an absorbing-state phase
transition~\cite{bak1987,dickman2000} tuned to a fixed critical
boundary by a slow drive balanced against fast relaxation.  As Gunji
\textit{et al.}~\cite{gunji2021} note for swarming criticality, such a
state ``is not self-organizing, and it requires parameter tuning.''
\BIB{} reaches criticality by a different route.  The inverse-Bayesian
step re-injects the least-supported hypothesis at the current empirical
distribution $\hat P_t$, redrawing the hypothesis support on the
simplex around the present pattern and pinning the per-step
log-posterior drift $d=\langle\eta\rangle$ to zero as a structural
identity of the update (Appendix~\ref{app:heuristic}) rather than as a
tuned set point.  In Gunji's natural-born-intelligence
framework~\cite{gunji2026}, inverse Bayesian inference is exactly this
boundary transformation, which places \BIB{} with the tuning-free,
reconstruction-based routes to criticality exemplified by
asynchronously-tuned cellular automata~\cite{gunji2014}.  Criticality
is thus a generic property of boundary reconstruction over a natural
parameter range rather than a fine point reached by self-organization
toward an absorbing state.  We therefore do not certify the mechanism
as SOC in the strict (absorbing-state) sense.  What the data establish
is the boundary-reconstructed critical structure, the design- and
$\Nh$-invariant exponent, which stands on its own.

% =============================================================================
\section{Robustness of the BIB universality}\label{sec:robust}

\subsection{Window-size and game-dimension dependence}\label{sec:robust-mN}\label{sec:results-robust}

We next probe the robustness of the BIB collapse along two control
axes, the history-window size $m$ and the game size $N$.
Figure~\ref{fig:robust}(a) shows the \BIB-\BIB{} argmax CCDFs as the
history window $m$ varies, pooling across the four designs in a
single panel so that the $m$ dependence is visible at one glance
(complementing the design-collapse already established in
Fig.~\ref{fig:univ}).  At $m\in\{10,20\}$ the CCDF tail is short and
the fitted exponent takes various values.  At $m\in\{50,100\}$ the
pooled \BIB-\BIB{} CCDFs settle onto the canonical $\alpha\approx 1.43$
form, parallel to the dashed reference.  The $m=100$ per-design
exponents are $\{1.468,\,1.450,\,1.453,\,1.459\}$ with mean $1.458$.  The window value $m=50$ adopted as canonical in Ibuka
and Sasai~\cite{ibuka2024} corresponds to the smallest window at
which the universal BIB regime is fully established.

We then test whether the BIB universality extends to
higher-dimensional cyclic-dominance games by running large-scale
simulations for $N\in\{5,7\}$ across all four designs ($\Nh=10$,
$m=50$, same as the $N=3$ analysis).  Combined with $N=3$ data this
gives $4\times 2\times 3 = 24$ \BIB-\BIB{} measurements.

The $N=3$ \BIB-\BIB{} universality
(cross-design range $0.03$, mean $1.43$) is recovered cleanly and
extends essentially unchanged to higher dimension, with the \BIB-\BIB{}
exponent staying at $\alpha\approx 1.4$ (means $1.43$, $1.41$, $1.38$ at
$N=3,5,7$) and every one of the $24$ cells lying inside both the
\Levy{} regime and the central band $[1.3,1.6]$.  The mild drift of
the mean and the small growth of the cross-design SD with $N$ ($0.016$,
$0.034$, $0.039$ at $N=3,5,7$) reflects the
larger fluctuations expected at higher game dimension rather than a
breakdown of the class.  \BO-\BO{}, by
contrast, does not collapse onto a single power law at any $N$
(Fig.~\ref{fig:robust}b), so we assign it no exponent.  Its argmax
CCDFs decay markedly faster than the BIB heavy tails.

Under a uniform truncated-power-law fit the heavy-tail signature is
preserved, with no broadening, as the game dimension grows.
Figure~\ref{fig:robust}(b) shows the underlying argmax-persistence
CCDFs directly, where the \BIB{} tails keep a common slope across $N$
while the \BO{} tails do not.

\subsection{Generality across games via matching pennies}\label{sec:robust-mp}

The game-dimension sweep above stays within the cyclic-dominance
family.  To test whether the BIB universality is intrinsic to the
inverse-Bayesian inference rule rather than to the cyclic structure of
RPS, we applied the identical inference core to a structurally
different zero-sum game with a uniform Nash equilibrium and no cyclic
dominance, matching pennies ($N=2$).  Each agent keeps the same
$\Nh=10$ hypotheses, the same Bayesian update with conditional
Jelinek--Mercer smoothing, and the same inverse-Bayesian renewal at
window $m=50$.  Only the observation alphabet (two symbols) and the
win/lose rule change (one agent matches, the other mismatches, and the
uniform Nash $(1/2,1/2)$ again makes any persistence internal).  The
setup is detailed in Appendix~\ref{app:matching}.

Both internal observables reproduce the on-off $3/2$ class.  Pooling over the two random-initialization
designs (sampling and argmax prediction), the argmax-persistence
exponent is $\alpha^{\mathrm{arg}}\approx 1.47$ ($1.467$ and $1.471$
for the two designs) and the laminar-phase exponent is
$\alpha^{\mathrm{lam}}\approx 1.42$ ($1.418$ and $1.419$), with a
truncated power law strongly preferred over an exponential in every
case (normalized log-likelihood ratio $R=75$ to $89$).  These values
coincide with the RPS exponents ($\alpha^{\mathrm{arg}}\approx 1.43$,
$\alpha^{\mathrm{lam}}\approx 1.325$).  The critical exponents are a
property of the inverse-Bayesian rule, not of the cyclic structure of
RPS.

This control also sharpens the finite-window reading of the residual
shift from $3/2$ discussed in Sec.~\ref{sec:soc-onoff}.  The
inverse-Bayesian step re-injects the renewed hypothesis near the
empirical histogram at an expected divergence that scales as
$(N-1)/(2m)$ (Appendix~\ref{app:heuristic}), so drift cancellation
becomes exact only as this residual vanishes.  Reducing the game size
from $N=3$ to $N=2$ halves the residual at fixed $m$, and accordingly
the matching-pennies argmax exponent ($\alpha\approx 1.47$) sits closer
to the asymptotic $3/2$ than its RPS counterpart ($\alpha\approx
1.43$), in the direction the finite-$m$ reading predicts.

\subsection{Robustness to observation rule}\label{sec:robust-obs}\label{sec:results-obsrule}

A further robustness axis tests whether the BIB universality is
specific to the observation rule of Eq.~(\ref{eq:reward}) or instead
reflects a structural feature of the renewal dynamics.  The reference rule makes two design
choices that are not forced by the BIB formalism: (i) skip the
posterior update on a tie, and (ii) sample a counterfactual
observation from $\mathcal{D}\setminus\{d_A\}$ on a defeat.  We run a
$2\times 2$ ablation over the tie-step and defeat-step branches of
Eq.~(\ref{eq:reward}), giving four schemes:
eq3 $=$ (skip, $\mathcal{D}\setminus\{d_A\}$), the reference
rule;
caseA $=$ (uniform, $\mathcal{D}\setminus\{d_A\}$), tie
outcome injects a uniform-noise observation;
hybrid $=$ (uniform, opponent), tie-uniform plus
ground-truth defeat observation;
defeatGT $=$ (skip, opponent), a ground-truth (GT) defeat
observation without compensating tie-uniform.
The first three schemes keep the inverse-Bayesian renewal active.  The
fourth suppresses it.

We ran the four schemes at medium scale
($T=10^{4}\times\,200$ runs) for all $4$ designs $\times$
\{\BIB-\BIB, \BO-\BO\} $=32$ cells, and verified the three renewal-preserving
schemes at large scale ($T=2\times 10^{5}\times\,20$ runs,
\BIB-\BIB{} only, $12$ cells).  For the argmax-persistence exponent
the three renewal-preserving schemes converge to a common value, with
cross-scheme means $\alpha(\Targmax)\in\{1.431,\,1.443,\,1.446\}$,
a maximum mean-to-mean difference of $0.015$, smaller than the
within-scheme cross-design spread of $0.016$--$0.030$
(Fig.~\ref{fig:scheme-ablation}, left).  A nonparametric bootstrap
($B=200$ truncated-power-law refits per cell) confirms that this
agreement is well resolved, with the per-cell interquartile range of
$\alpha(\Targmax)$ at $\le 0.020$ for all $12$ cells (an order of
magnitude below the differences being compared) and every cell
remaining in the \Levy{} regime $1<\alpha\le 3$.

The residual offsets between renewal-preserving schemes are, however,
observable dependent.  Computed under a single fixed fitting
convention (forced $x_{\min}=1$, threshold $\theta=0.4$), the laminar
exponent of the most aggressive renewal-preserving scheme (hybrid) sits slightly
\emph{above} that of the reference rule, at
$\alpha^{\mathrm{lam}}_{\mathrm{eq3}}=1.41\pm0.003$ versus
$\alpha^{\mathrm{lam}}_{\mathrm{hybrid}}=1.48\pm0.002$ (cross-design
SD, $4$ designs), a displacement of $\approx +0.07$ that is small in
absolute terms yet large relative to the within-scheme cross-design
scatter ($\approx 0.003$)~\footnote{The absolute laminar exponent is
sensitive to the power-law fitting convention.  The default $x_{\min}$
scan and different library versions shift it by $\sim 0.07$, so the
cross-scheme laminar comparison is reported here under a single fixed
$x_{\min}=1$ convention.  The headline value $\alpha^{\mathrm{lam}}=
1.325$ quoted in Sec.~\ref{sec:soc-onoff} was obtained with the
default scan.  Recomputed under the
fixed convention used here the reference rule gives $1.41$, consistent
with the comparison above.}.  The same upward, design-independent
offset appears in the posterior-sharpness scaling
$\sigma(\Ph)\sim\Nh^{-\beta}$, where hybrid yields
$\beta=1.119\pm0.011$ against the reference $1.067\pm0.008$, and in the
central-regime hand-number scaling ($\Nh\in\{6,10\}$), where hybrid's
$\alpha(\Targmax)$ tracks the reference values up to the same
$\approx +0.02$ shift.  The three renewal-preserving schemes therefore do not
collapse onto a single point in exponent space but onto a tightly
clustered family, with each scheme being cross-design universal and
the small ($\lesssim +0.07$) systematic offsets between schemes
reflecting their slightly different renewal rates rather than any
breakdown of universality.

The defeatGT scheme,
by contrast, injects deterministic ground-truth feedback on every
defeat step without compensating stochastic updates on ties, shifting
the attractor to a distinct, steeper power-law exponent
$\alpha\approx 2.31$ (truncated power law preferred, $\sim\!2.3$ decades,
and unlike \BO{} this passes the heavy-tail criterion) with sharper
posteriors
($\sigma(\Ph)\approx 0.09$ versus $\approx 0.13$ in the renewal-preserving
schemes) and Nash-deviating macro-rates (win/tie/defeat $\approx
0.28/0.44/0.28$ versus $\approx 0.33/0.33/0.33$ in the renewal-preserving schemes).  This
shows that the critical
regime is selected by whether the inverse-Bayesian renewal remains
operative, not by the specific observation rule.  Rules that keep the
renewal active remain cross-design universal and cluster within
$\lesssim 0.07$ in exponent space, while a rule that suppresses it
departs from the universal attractor entirely.

% -----------------------------------------------------------------------------

% =============================================================================
\subsection{Robustness to likelihood sharpness}\label{sec:robust-sharp}

A third control axis is the sharpness
$\alpha_{\mathrm{init}}\in\{0.4,\dots,0.9\}$ of the structured $\Pdh$
templates.  The sweep is restricted to structured-initialization
designs (ss, sa), because random initialization does not admit a
controllable peak.  Every one of the $n=12$ conditions stays in the
\Levy{} regime $1<\alpha\le3$ on both observables, the
truncated-power-law form being preferred in $11/12$ (argmax) and
$10/12$ (laminar).  The \BIB-\BIB{} argmax exponent holds a tight band
$\alpha^{\mathrm{arg}}=1.48\pm0.02$ ($1.44$--$1.51$), with only a weak
upward drift ($\approx+0.04$ from $\alpha_{\mathrm{init}}=0.4$ to $0.9$).  It sits slightly above the pooled four-design value $1.43$ because the
structured designs occupy the upper end of the design spread
(Sec.~\ref{sec:univclass}).  The laminar exponent shows no sharpness
trend, $\alpha^{\mathrm{lam}}=1.34\pm0.02$ ($1.30$--$1.37$), consistent
with its canonical $1.325$.  \BO-\BO{} forms no comparable power law
over the same sweep (Appendix~\ref{app:modelselect}).  The class
membership is thus robust to sharpness, both observables staying in the
$3/2$ class across the sweep, the argmax with only a weak drift and the
laminar with none.  Combined with the design- and $\Nh$-invariance of
Sec.~\ref{sec:univclass} and Sec.~\ref{sec:fss-nhinv}, the BIB
universality holds across three independent parameter axes:
implementation design, hypothesis count $\Nh$, and likelihood sharpness
$\alpha_{\mathrm{init}}$.

\begin{table}[htbp]
\centering
\small
\caption{$\Pdh$-sharpness sweep, truncated-power-law exponent
statistics over $\alpha_{\mathrm{init}}\in\{0.4,\dots,0.9\}$ ($6$
values) for the two structured designs (ss, sa) and both agents
(mean $\pm$ SD over the $n=12$ design\,$\times$\,sharpness
conditions).  \BIB{} stays close to its canonical exponents on both
observables (SD $\approx 0.02$).  \BO{} under the same model selection forms
no comparable power law and is omitted.}
\label{tab:sharpness}
\begin{ruledtabular}
\begin{tabular}{llcc}
Observable & Scheme & mean $\pm$ SD & $\alpha$ range \\
\hline
Argmax persistence & \BIB & $1.48 \pm 0.02$ & $1.44$--$1.51$ \\
Laminar phase      & \BIB & $1.34 \pm 0.02$ & $1.30$--$1.37$ \\
\end{tabular}
\end{ruledtabular}
\end{table}

\section{Reward analysis of internal-state versus external functionality}\label{sec:reward}\label{sec:results-reward}

\subsection{Direct comparison of BIB and BO win rates}\label{sec:reward-direct}

\begin{figure*}[tp]
\centering
\includegraphics[width=\linewidth]{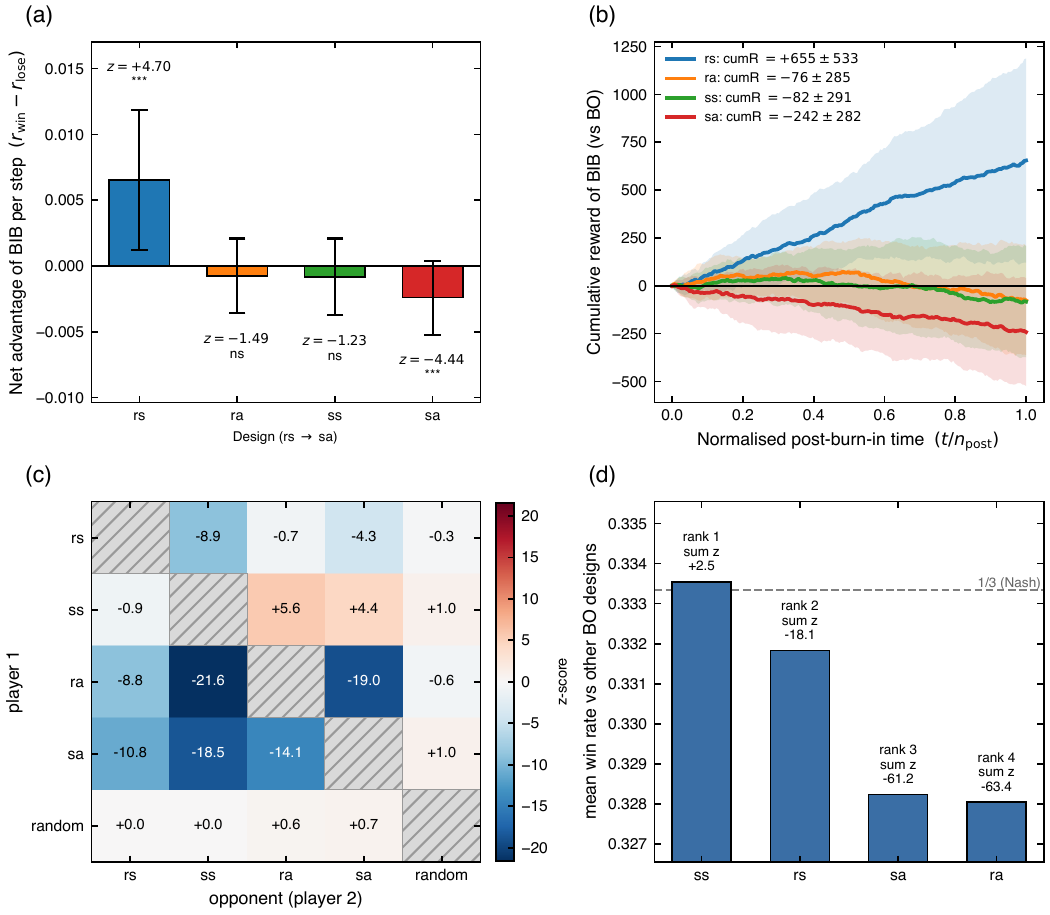}
\caption{\BIB{} versus \BO, head-to-head reward and cross-design
tournament (at $m=50$, $\Nh=10$).  (a)~Net advantage of \BIB{} per
step $\langle r_{\mathrm{win}}-r_{\mathrm{defeat}}\rangle$ by design,
with the win-rate $z$-score against the Nash value $1/3$ annotated
($n=20$ runs).  (b)~Cumulative reward of \BIB{} over the post-burn-in
second half of each run (mean $\pm$ SD over $20$ runs, final value in
the legend).
(c)~$5\times5$ cross-design tournament, the $z$-score of player~1's win
rate against $1/3$ for every pairing (bold entries are significant at
$|z|>1.96$, hatched diagonal cells are self-matches).  (d)~\BO-internal
tournament ranking within the \BO{} field (excluding the Nash
baseline), giving ss~$>$~rs~$>$~sa~$>$~ra.}
\label{fig:bib-vs-bo}
\end{figure*}

Although the persistence-time universality is robust to design,
the actual win-rate balance between \BIB{} and \BO{} in head-to-head
play depends strongly on design
(Fig.~\ref{fig:bib-vs-bo}a,b).  The per-design win rates against the
Nash value $1/3$ are $\{0.3369, 0.3329, 0.3329, 0.3320\}$ for
(rs, ra, ss, sa), with $z=\{+4.70, -1.49, -1.23, -4.44\}$ ($n=20$ runs,
rs and sa significant at $p<0.001$).  The pattern is monotonic across
the four designs, running from rs (\BIB-favored) through the
near-Nash ss and ra ($|z|<1.5$) to sa (\BO-favored).

The BIB universality of $\alpha\approx 1.43$ exists at the
\emph{internal-state} level (argmax persistence within the agent's
hypothesis posterior) and does not directly translate into external
win-rate dominance.

\subsection{BO cross-design tournament with Nash baseline}\label{sec:reward-tournament}\label{sec:results-tournament}

Two alternative explanations of the monotonic BIB-vs-BO pattern of
Fig.~\ref{fig:bib-vs-bo}a need to be ruled out before this
internal-vs-external separation can be firmly established:
\emph{(i)}~the pattern might reflect an internal \BO{} strength ranking
that aligns with the design hardness axis, and
\emph{(ii)}~the \BO{} designs themselves might drift from Nash
equilibrium in a way that biases the head-to-head comparison.  To rule
out both, we ran a $5\times 5$ tournament, each of the four \BO{}
designs plus a Nash random baseline paired against itself and against
each other strategy ($T=200{,}000\times\,20$ runs per cell).

\paragraph{Conservation of Nash equilibrium.}
All four \BO{} designs achieve $z\in[-0.55,\,+1.00]$ against Nash
random, within the window expected from Nash-equilibrium theory
(the random row/column of Fig.~\ref{fig:bib-vs-bo}c).  The simulator and inference dynamics
conserve Nash equilibrium to within statistical precision, ruling
out alternative \emph{(ii)}.

\paragraph{BO-internal ranking.}
Within the BO field (excluding the Nash random baseline), the four
designs rank as ss~$>$~rs~$>$~sa~$>$~ra (Fig.~\ref{fig:bib-vs-bo}d),
with the argmax-mode designs (ra, sa)
losing to all opponents in their row while the sample-mode designs
(rs, ss) win or tie against most opponents, consistent with the
multi-armed-bandit literature, in which posterior sampling outperforms
greedy argmax in regret-bounded
settings~\cite{agrawal2011,russo2014,odonoghue2021}.

\paragraph{BO strength and the BIB advantage.}
The BIB-vs-BO ordering (rs $\succ$ ss $\succ$ ra $\succ$ sa,
Fig.~\ref{fig:bib-vs-bo}a) and the
BO-internal strength ordering (ss $\succ$ rs $\succ$ sa $\succ$ ra,
Fig.~\ref{fig:bib-vs-bo}d) do
not coincide, with the design that is strongest within the \BO{} field
(ss) being in fact BIB-\emph{disfavored} ($z=-1.23$) while the most
BIB-favored design (rs, $z=+4.70$) ranks only second in \BO{}
strength.  The BIB advantage is non-monotonic in \BO{}
strength, since even the weakest \BO{} design (ra) is only a tie.  The
head-to-head pattern therefore
cannot be a function of \BO{} strength (the two orderings share no
common rank position, and their Spearman correlation, $\rho=0.60$ at
$n=4$, is not significant).  This rules out alternative \emph{(i)}.

\paragraph{Internal versus external functionality.}
Together, the reward analysis and the tournament thus
clarify that the BIB universality is an \emph{internal-state}
property.  When a BIB agent plays head-to-head against a BO agent, the
direction of the win-rate advantage depends systematically on the
shared design (BIB beats BO at the rs corner with $z=+4.70$, loses
at the sa corner with $z=-4.44$, ties in between), even though the
BIB exponent itself is unchanged across designs, and the BIB
win-rate pattern is not reducible to opponent strength.  The
universality of $\alpha\approx 1.43$ is therefore a signature of the
internal hypothesis-renewal dynamics, not of overt game performance.
This separation has direct implications for the comparison between
BIB-style generative agents and Nash-targeting algorithms such as the
CFR family~\cite{zinkevich2007,tammelin2014,brown2019} and recent
LLM-based agents~\cite{zheng2025}.  In symmetric zero-sum RPS, all
CFR-family algorithms reduce to uniform random
play~\cite{neller2013}, and LLM agents reproduce human-like deviations
descriptively but apply them rigidly~\cite{zheng2025}.  Neither
generates the parameter-invariant heavy-tailed signature we observe
in BIB.  Unlike the superhuman optimizers and language-model agents
above, which reach their play only through extensive self-play or
large-scale pre-training, BIB generates this signature online,
directly from the inference rule and with no pre-training.

% =============================================================================
\section{Behavioral implications, limitations, and outlook}\label{sec:behaviour}\label{sec:disc-behaviour-outlook}

We take BIB inference as a candidate generative process for the
heavy-tailed persistence statistics reported in human and animal
decision-making across diverse
substrates~\cite{barabasi2005,vazquez2005,karsai2012,wang2014,arai2025}.

\subsection{Behavioral scope of the mechanism}\label{sec:scope}

\begin{figure*}[tp]
\centering
\includegraphics[width=\linewidth]{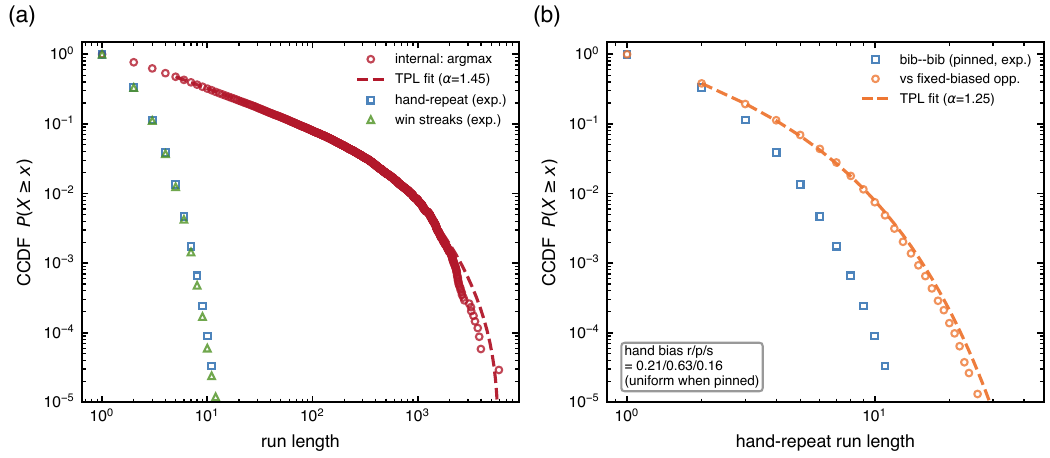}
\caption{Internal-state criticality versus behavioral
expression.  Complementary cumulative distributions (CCDFs),
\BIB--\BIB{}, $\Nh=10$, stationary window.  (a)~At the uniform
Nash fixed point the heavy tail is internal only.  Argmax persistence is
a power law (truncated-power-law fit $\alpha\approx1.45$, dashed
guide), while
behavioral hand-repeat runs and win streaks are exponential.
(b)~The same behavioral observable (hand-repeat runs) once
the adversarial pinning is removed.  Against a fixed biased opponent
$(0.6,0.2,0.2)$ the agent locks onto an exploiting hypothesis,
$P(d\mid h^{*})$ concentrates, the action marginal becomes biased
(hand frequencies $0.21/0.63/0.16$ versus uniform when pinned), and the
hand-run tail crosses over from exponential to a power law
($\alpha\approx1.25$).}
\label{fig:scope}
\end{figure*}

The internal-state nature of the universality has a sharp behavioral
corollary that delimits the scope of the BIB mechanism.  In symmetric
\BIB--\BIB{} play at the uniform Nash fixed point, the heavy-tailed
persistence is confined to the internal hypothesis dynamics, where the
argmax-persistence distribution is a power law
($\alpha\approx 1.43$) whereas every behavioral streak statistic
(hand-repeat runs, win, win-or-draw, and defeat streaks) is
exponential (Fig.~\ref{fig:scope}(a)).  The mechanism is the
cyclic-dominance structure of the game, in which a symmetric adaptive
opponent best-responds to and thereby competes away any exploitable
bias, so
the action marginal is pinned to uniform Nash and the internal
critical fluctuations cannot imprint on the symbol stream.  Each hand
is, at equilibrium, a near-uniform sample from $P(d\mid h)$ regardless
of which hypothesis is currently held, a high-entropy read-out that
erases the temporal correlation carried by the internal state.  In
this regime BIB is behaviorally indistinguishable from a
Nash-targeting solver of the CFR
family~\cite{zinkevich2007,tammelin2014,brown2019} (both reduce to
uniform random play~\cite{neller2013}), even though BIB's internal
dynamics are critical and the solver's are not.

Crucially, this pinning is a property of the adversarial symmetry of
the equilibrium, not of BIB itself.  When the opponent carries an
exploitable structure that it does not adapt away, which is the
situation a fixed or slowly-varying non-Nash player presents, the
latent flexibility is expressed.  Figure~\ref{fig:scope}(b) shows that
against a fixed biased opponent the agent's action marginal becomes
strongly biased and the behavioral hand-run distribution crosses over
from exponential to a power-law tail ($\alpha\approx 1.25$).  The
internal critical state thus acts as a
reservoir of behavioral flexibility that is gated by the opponent's
exploitability rather than expended at equilibrium.  This is the sense
in which the scope of BIB differs from that of Nash-targeting
algorithms.  The CFR family converges to uniform play and remains
there, optimal against a worst-case adversary but rigid against a
structured one, possessing neither internal critical dynamics nor a
route to flexible exploitation.  BIB attains the same Nash-safe
behavior at equilibrium while retaining a critical internal state that
tracks and exploits non-Nash structure.  Human opponents, with their
heavy-tailed action timing and systematic
biases~\cite{barabasi2005,dong2014,xuharvey2014,spanknebel2015},
inhabit precisely this structured, exploitable regime.  The present
substrate therefore yields a concrete, falsifiable prediction for
human-versus-BIB play, and identifies the observables that test it
(argmax persistence internally, and hand-run, win-streak, or transition-run tails
behaviorally, as appropriate to the opponent's structure).  The persistence statistics that are invisible in
symmetric self-play should re-emerge against human opponents as BIB
tracks and exploits their deviations from Nash.  A re-analysis of
existing human-versus-bot play already exhibits the predicted
behavioral crossover (Sec.~\ref{sec:human}), an encouraging if
indirect first test.

\subsection{Behavioral signature in existing human play}\label{sec:human}

The prediction above is already partially borne out by existing data.
The behavioral corollary predicts that a player's persistence is
short-tailed against an adaptive opponent, which pins the action
stream toward the uniform Nash point, and crosses over to a heavy
tail against an exploitable opponent, onto which the agent locks.  We
tested this on the public human-versus-bot data of Brockbank and
Vul~\cite{brockbank2024}, pooling their two experiments: $300$-round
games against seven fixed, stationary-pattern bots (exploitable by the
participant) and against adaptive bots that instead exploit the
participant.  Ordering the fifteen bot conditions by a single empirical
axis of exploitability, the human win rate, and applying a
$\ge 50$-round inclusion filter, the pool comprises $n=451$ completed
games ($244$ fixed, $207$ adaptive).  As these bots are
transition-based, we analyze the persistence of the player's transition
sequence.

\begin{figure*}[tp]
\centering
\includegraphics[width=\linewidth]{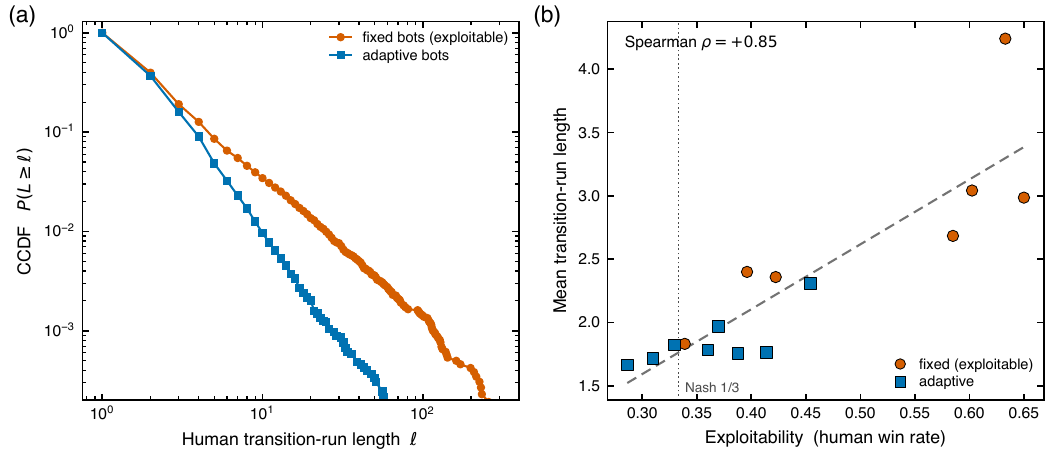}
\caption{The predicted behavioral crossover in existing human play.
Re-analysis of the public human-versus-bot
data of Brockbank and Vul~\cite{brockbank2024} ($n=451$ games across
fifteen bot conditions in two experiments).  (a)~CCDF of the human
transition-run length, the tail being markedly heavier against fixed,
exploitable bots than against adaptive bots, as predicted for
exploitable versus equilibrium-pinning opponents.  (b)~Across the
fifteen conditions, the mean transition-run length increases
monotonically with empirical exploitability (human win rate, Spearman
$\rho=+0.85$, $p=10^{-4}$), with a truncated power law preferred over an
exponential in every condition.}
\label{fig:human}
\end{figure*}

Both behavioral predictions hold [Fig.~\ref{fig:human}].  The
transition-run tail is markedly heavier against the fixed than against
the adaptive bots, and across the fifteen conditions the mean
transition-run length increases monotonically with empirical
exploitability (Spearman $\rho=+0.85$, $p=10^{-4}$), with a truncated
power law preferred over an exponential in every condition.  This
upgrades the human-play link from a prediction to preliminary empirical
support.  It remains consistent with, but does not by itself establish,
the internal criticality of Sec.~\ref{sec:soc}: the argmax-persistence
statistic is a \emph{latent} property of the inference, not directly
observable in the human symbol stream.  A direct test, fitting \BIB{}
and control learners to individual human play, is the aim of the
planned human-versus-agent experiments.  The data, conditions, and fits
are detailed in Appendix~\ref{app:human}.

\subsection{Alternative explanations, limitations, and outlook}\label{sec:limitations}

\paragraph{Alternative explanations considered.}
Four non-mechanistic readings of the BIB universality were examined
and found insufficient.  \emph{(i)}~fit-function arbitrariness is ruled
out, since the exponent is chosen by Akaike weights over power-law,
truncated-power-law, and exponential candidates
(Sec.~\ref{sec:methods-sweep}), and the central claims rest not on the
bare value $1.43$ but on the cross-design \emph{collapse} and
$\Nh$-invariance, which are insensitive to the $x_{\min}$ convention.
\emph{(ii)}~the conditional-smoothing regularizer is only an
implementation detail that prevents posterior collapse, and
Appendix~\ref{app:smoothing} shows the core results survive when it is
changed.  \emph{(iii)}~the prediction mode is not the source, since
both argmax- and sample-mode designs are among the four that collapse
onto the common exponent.
\emph{(iv)}~the BO design-dependence is not a mere implementation
difference.  The cross-design tournament (Sec.~\ref{sec:results-tournament})
shows that the BO-internal strength ordering does not match the
BIB-vs-BO advantage ordering, and BO, which lacks the inverse step,
serves as a \emph{negative control} that exhibits no analogous
universality.  The BIB/BO distinction is therefore one of renewal
mechanism, not of raw performance.

Several limitations bound the present results.  We emphasize that we
claim \emph{evidence} for internal-state SOC-\emph{like} criticality
rather than a proof of classical SOC, and that $\alpha\approx 1.43$ is
a protocol-dependent estimate of a universality class under the tested
sweep, not a physical constant.  Indeed, from an external,
observational standpoint the self-organized-criticality label is
itself fragile, since the ``self-organization'' it invokes can conceal
an implicit tuning, so whether a system counts as SOC depends on the
criteria one imposes.  Our aim is therefore not to certify the
mechanism as self-organized criticality per se, but to characterize
the universal structure of the criticality
itself~\cite{gunji2021,mora2011,munoz2018}, namely the design-
and $\Nh$-invariant exponent, which stands independently of the SOC
label.
\emph{(i)}~Human data: although a re-analysis of existing
human-versus-bot play already exhibits the predicted behavioral
crossover (Sec.~\ref{sec:human}), we have not yet directly compared
BIB-generated persistence statistics with human gameplay at the level
of the latent internal observable.
\emph{(ii)}~Multi-player extension: the two-player setting is the
minimal nontrivial case.  The population-level extension with explicit
replicator dynamics is not treated here.
\emph{(iii)}~Time-varying game rules: settings in which the
cyclic-dominance order itself changes during play, such as a cycle
reversal, could test, and thereby differentiate, BIB's
flexible internal critical dynamics.
\emph{(iv)}~Formal SOC criteria beyond FSS: while empirical
finite-size scaling is established (Sec.~\ref{sec:results-fss}), the
explicit construction of avalanche-style hypothesis-renewal events
with scale-invariant size distributions, and a first-principles
derivation of $z$ together with the relation between $z$ and the
laminar-phase exponent, lie beyond the present scope.
\emph{(v)}~Update-rule variants: the present paper characterizes the
\emph{argmin-replacement} update rule~\cite{gunji2018,gunji2021,ibuka2024}.
Whether the \emph{argmax-modification} rule of Shinohara
\textit{et al.}~\cite{shinohara2020,shinohara2021,shinohara2022}
shares the same universality class is not yet established.
\emph{(vi)}~Observation-space locality: our defeat-time observation
rule implicitly references the global hand set.  Whether
locally-scoped observation rules sharpen the universality at
higher $N$ is an open question.

% =============================================================================

% =============================================================================
\section{Conclusion}\label{sec:conclusion}

We have characterized the empirical universality class of
Bayesian--inverse-Bayesian (BIB) inference in the minimal discrete
decision-making setting of $N$-hand rock-paper-scissors, using
large-scale simulations across a six-axis parameter space ($\sim\!10^8$
decision events in total).  Across the four implementation designs and three opponent types, \BIB{}
remains in the same internal critical state, its argmax-persistence
distribution a power law with exponent $\alpha = 1.43\pm 0.02$ at the
canonical window $m=50$.  Along the window axis the exponent is not constant but
rises monotonically toward the on-off value $3/2$ as $m$ grows (mean
$1.458$ at $m=100$), the systematic offset being fixed by the
finite-sample residual $(N-1)/(2m)$.  Under hypothesis-count rescaling, the same exponent is
invariant in the core central regime $\Nh\in\{6,10\}$
(cross-design mean $1.433$, SD $0.014$, $n=8$) even while Bayes-only
inference, lacking the renewal step, produces no comparable power law and
hence no universal regime.  The \BIB-\BIB{}
laminar-phase length distribution exhibits the truncated power law
$\alpha^{\mathrm{lam}} = 1.325\pm 0.006$, lying within $12\%$ of the
universal value $3/2$ of on-off
intermittency~\cite{platt1993,heagy1994}, and the cutoff time follows a
finite-size scaling $T_{\max}(\Nh)\propto \Nh^{-z}$ with
$z = 1.49\pm 0.07$, consistent with the critical-state ansatz
Eq.~(\ref{eq:fss}).

Taken together, these findings identify BIB inference as a generator
of parameter-invariant heavy-tailed persistence statistics, produced
by a boundary-reconstructing renewal mechanism on the internal
hypothesis-space simplex.  The argmax persistence time of any single
hypothesis is dual to the timescale of hypothesis-space
reorganization.  The inverse-Bayesian renewal (relaxation) that
periodically reseeds the hypothesis set renders the dynamics critical
without external parameter adjustment.  The
spread-scaling exponent $\beta_{\BIB}=1.067\pm 0.008$, close to the
information-theoretic lower bound $\beta=1$, is consistent with the
inverse step continually returning the posterior to a near-equilibrium
state.  We read the mechanism as a boundary-reconstructing route to
internal-state criticality, complementary to self-organized
criticality, and identify it operationally with on-off intermittency
in hypothesis space, with the small systematic shift from $3/2$
admitting a natural reading as the extended on-off intermittency
regime of Bertin~\cite{bertin2011}.

The internal-state versus external-functionality separation revealed by
the reward analysis shows that BIB's exponent is a
signature of the agent's internal renewal dynamics, not of overt game
performance against a non-renewing opponent.  The most direct next step is to confront the same six-axis protocol
with human-gameplay data, testing whether natural decision-making
reproduces the heavy-tailed persistence signature identified here.
In short, our contribution is not a claim that BIB realizes
self-organized criticality, but the identification of a universal
critical structure, namely the design- and $\Nh$-invariant exponent,
generated by adding inverse-Bayesian relaxation (hypothesis renewal)
to Bayesian inference and standing independently of the SOC label.
This mechanism thereby offers a candidate route by which natural
gameplay may be rendered as a generative process rather than a
descriptive imitation~\cite{gunji2026}.

% =============================================================================
%  End-matter
% =============================================================================

\paragraph*{Author contributions.}
Following the CRediT (Contributor Roles Taxonomy) standard.
\textbf{K.~S.} (Kazuto Sasai): Conceptualization, Methodology,
Software, Formal analysis, Investigation, Data curation, Writing --
original draft, Writing -- review \& editing, Visualization,
Supervision, Project administration, Funding acquisition.
\textbf{Y.-P.~G.} (Yukio-Pegio Gunji): Conceptualization (BIB
framework), Methodology (theoretical
positioning), Writing -- review \& editing.
Both authors have read and agreed to the submitted version of the
manuscript.

\paragraph*{Competing interests.}
The authors declare no competing interests.

\paragraph*{Funding.}
This work was supported by the Japan Society for the Promotion of
Science (JSPS) KAKENHI grant numbers~22K12143 and 26H01196.

\paragraph*{Data availability.}
The data that support the findings of this article are publicly
available~\cite{sasai2026data}.  The data were generated by numerical
simulations, and the simulation source code, intermediate results, and
figure-generation scripts that reproduce every figure and numerical
claim are publicly available in the same archive~\cite{sasai2026data}
and in the repository
\url{https://github.com/kazsasai/bayesian-inverse-bayesian-rps}.

\paragraph*{Declaration of AI use.}
The authors used Anthropic Claude (Claude Opus~4.7, 4.8) as a
research-assistance tool for
(i) verifying numerical consistency between the LaTeX manuscript and
the simulation data, (ii) re-generating all production figures from
the raw output of the simulator, and (iii) drafting and reformatting
sections of this manuscript according to author guidance.  All
scientific content, design decisions, interpretations, and
conclusions are the authors' own.  The assistant did not generate any
factual statement that was not separately verified against the
simulation outputs by the authors.

% =============================================================================
%  Appendices
% =============================================================================
\appendix

\section{Conditional smoothing as a relaxation mechanism}\label{app:smoothing}

Section~\ref{sec:methods-obs} identifies the posterior regularization
$P(h)\leftarrow 0.91\,P(h)+0.015$ if $\min P(h)<0.002$ as a
\emph{conditional} Jelinek--Mercer interpolation~\cite{jelinek1980,
chen1996} with $\Nh$-dependent strength
$\alpha_{\mathrm{JM}}(\Nh)\approx 0.14$ at $\Nh=10$.  We argued there
that the \emph{conditional} aspect of this rule, which applies only when
posterior collapse threatens, is essential to the BIB dynamics.
This appendix verifies that claim experimentally through a controlled
4-way comparison of regularization rules.

\subsection{Comparison design}
We ran a controlled comparison at the canonical condition (rs design,
\BIB-\BIB{}, $\Nh=10$, $m=50$, $T=2000$ steps with the last $1000$
analyzed, $100$ runs, identical seeds across implementations).  Each
implementation differs only in the posterior-regularization rule
(Table~\ref{tab:smoothing-impl}).

\begin{table*}[tp]
\centering
\small
\caption{Four smoothing implementations compared in this appendix.}
\label{tab:smoothing-impl}
\begin{ruledtabular}
\begin{tabular}{ll}
Implementation & Description \\
\hline
(1) Ibuka exact &
  \parbox[t]{0.62\textwidth}{\raggedright
  $P\leftarrow 0.91 P+0.015$ if $\min P<0.002$, then
  renormalize.  Reference dynamics.} \\[2pt]
(2) Cond.\ JM $\alpha_{\mathrm{JM}}=0.15$ &
  \parbox[t]{0.62\textwidth}{\raggedright
  $P\leftarrow 0.85 P+0.15/\Nh$ if $\min P<0.002$.
  Equivalent to (1) up to a small parameter change.} \\[2pt]
(3) Cond.\ JM $\alpha_{\mathrm{JM}}=0.142$ &
  \parbox[t]{0.62\textwidth}{\raggedright
  Same as (2) with $\alpha_{\mathrm{JM}}$ chosen to match (1)
  at $\Nh=10$.} \\[2pt]
(4) Always-on JM $\alpha_{\mathrm{JM}}=0.10$ &
  \parbox[t]{0.62\textwidth}{\raggedright
  $P\leftarrow 0.9 P+0.10/\Nh$ at every step (no trigger).} \\
\end{tabular}
\end{ruledtabular}
\end{table*}

\subsection{Dynamics of the four rules}
\begin{figure}[tbp]
\centering
\includegraphics[width=\linewidth]{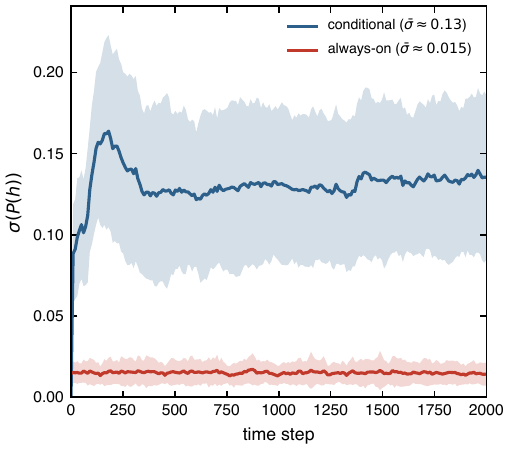}
\caption{$\sigma(\Ph)$ dynamics for conditional (blue) and always-on
Jelinek--Mercer (red) posterior smoothing (bib--bib self-play,
$\Nh=10$, $T=2000$, mean $\pm$ SD over $100$ runs).  A single
representative conditional rule is shown, the three conditional
implementations of Table~\ref{tab:s6} being statistically
indistinguishable.  Per-implementation $\bar\sigma$ values are in
Table~\ref{tab:s6}.}
\label{fig:smoothing}
\end{figure}

The three conditional rules are statistically indistinguishable in
their dynamics ($\bar\sigma\approx 0.13$, argmax-persistence
$\alpha\approx 1.36$--$1.38$, mean argmax run length $\approx 34$
steps, similar inter-run SD).  The always-on rule shows an $\sim 8$-fold
collapse in $\bar\sigma$, an $\sim 8$-fold increase in argmax switches,
a collapse of the mean argmax run length from $\approx 34$ to
$\approx 4$ steps, and a $\sim 25$-fold reduction in inter-run SD
(Table~\ref{tab:s6}, Fig.~\ref{fig:smoothing}).
The always-on truncated-power-law exponent is \emph{not} steeper but if
anything slightly shallower ($\alpha\approx 1.26$ versus $1.36$).  The
suppression of the heavy-tailed persistence is therefore one of
\emph{scale}, not slope.  The persistence \emph{cutoff} collapses
(the mean run length falls from $\approx 34$ to $\approx 4$ steps, an
$\sim 8$-fold rise in switches), matching the near-exponential always-on limit
of the reduced map in Appendix~\ref{app:heuristic}.

\begin{table}[htbp]
\centering
\small
\caption{Smoothing-implementation comparison at the canonical \BIB-\BIB{}
rs condition, with implementations (1)--(4) defined in
Table~\ref{tab:smoothing-impl}.  The three conditional rules are
statistically indistinguishable in $\bar\sigma$.  The always-on rule
shows an $\sim 8$-fold collapse in $\bar\sigma$, an $\sim 8$-fold
increase in argmax switches (mean run length $\approx 34\to 4$ steps),
and a $\sim 25$-fold reduction in inter-run SD.}
\label{tab:s6}
\begin{ruledtabular}
\begin{tabular}{crrlr}
Impl. & $\alpha$ & $\bar\sigma$ & best fit & $n_T$ \\
\hline
(1) & $1.358$ & $\mathbf{0.133\pm 0.027}$ & tpl & $5{,}787$ \\
(2) & $1.379$ & $0.131\pm 0.028$ & tpl & $5{,}897$ \\
(3) & $1.379$ & $0.133\pm 0.028$ & tpl & $5{,}951$ \\
(4) & $1.257$ & $\mathbf{0.016\pm 0.001}$ & tpl & $46{,}205$ \\
\end{tabular}
\end{ruledtabular}
\end{table}

\subsection{Why conditional triggering matters}
The conditional structure itself is essential.  The conditional
trigger acts as a \emph{relaxation mechanism} that fires only when
the Bayesian concentration brings the system to the edge
of collapse, mirroring the on-off intermittency mechanism of Platt,
Spiegel and Tresser~\cite{platt1993} in which a slow driving variable
intermittently triggers fast bursts.  An always-on regularization, by
contrast, removes the slow build-up phase altogether, clamping the
driving variable below the bursting threshold at every step in the
language of Platt, Spiegel and Tresser~\cite{platt1993}, so no laminar
build-up can accumulate and the on-off (laminar/burst) alternation,
and with it the heavy-tailed persistence signature, is suppressed.

% -----------------------------------------------------------------------------

\section{Methodological details}\label{app:methods}

This appendix consolidates methodological details referenced from
Sec.~\ref{sec:model}.

\subsection{Simulation scale and runs}\label{app:scale}

The main results use $T=2\times 10^{5}$ steps and
$20$ independent runs per condition.  After discarding the first
$T/2=10^{5}$ steps as burn-in, the remaining
$10^{5}\times\,20=2\times 10^{6}$ decision events are pooled for
statistical analysis.  For the central design $\times$ window
$\times$ pair sweep ($96$ conditions) this yields
$\sim\!1.9\times 10^{8}$ decision events in total.  The
$\Nh$ sweep at $5\times 24=120$ large-scale conditions analyzes an
additional $\sim\!2.4\times 10^{8}$ decision events.

\subsection{Metrics}\label{app:metrics}

The principal observable is the \emph{argmax-hypothesis persistence
time} $\Targmax$, defined as the run-length of consecutive time
steps $t$ on which the argmax index
$i^{*}(t):=\arg\max_i P_t(h_i)$ remains unchanged.  Long persistence
indicates the agent continuously trusts a single dominant hypothesis.
Short persistence indicates rapid hypothesis switching.  For each of
the two agents we compute $\Targmax^{(A)}$ and $\Targmax^{(B)}$ and
pool them across runs.

For the on-off intermittency analysis we additionally use:
(i) the \emph{laminar phase length}
$\Tlam(\theta)$ = the run-length of consecutive steps with
$\max_h P(h)>\theta$, with $\theta=0.4$ adopted as canonical (well
above $1/\Nh=0.1$, well below typical sharp-peak values $\sim 0.7$);
and (ii) the \emph{plateau length} $\Tpl(\theta,f)$ = the run-length
of argmax-stable intervals for which $P(h^{*})>\theta$ for at least
a fraction $f$ of the run, with $\theta=0.3$, $f=0.8$ as canonical.

\subsection{Implementation and reproducibility}\label{app:impl}

All simulations are implemented in Python (NumPy and multiprocessing
for parallelization, \texttt{powerlaw} for fitting, pandas for
aggregation).  The full source code, simulation scripts,
JSON-serialized intermediate results, and aggregate CSVs are publicly
available at \url{https://github.com/kazsasai/bayesian-inverse-bayesian-rps}
and archived on Zenodo with DOI
\texttt{10.5281/zenodo.20813022}.

\subsection{$x_{\min}$ and fit-convention sensitivity}\label{app:xmin}

To confirm that the BIB universality is not an artifact of the
$x_{\min}$ choice, we refit the pooled argmax-persistence times
($T_{\mathrm{argmax}}$ of both agents, \BIB-\BIB, $m=50$, $N=3$) under
several conventions (Table~\ref{tab:xmin}).  The \emph{absolute}
exponent depends on the convention.  A truncated power law with the
Clauset auto-$x_{\min}$ (which selects $x_{\min}=5$ for all four
designs) gives $\alpha\approx 1.43$.  Forcing $x_{\min}=1$ lowers it to
$\approx 1.28$ by including the short-run contamination.  A pure
power law gives $\approx 1.59$.  The \emph{cross-design spread},
however, stays at $\le 0.04$ under every convention, so the design
collapse, and with it the $\Nh$-invariance reported in
Sec.~\ref{sec:results-nh}, is the robust result, not the bare
value.  Absolute exponents should therefore be compared only within a
single fixed convention, as done throughout the main text.

\begin{table}[htbp]
\centering
\small
\caption{Sensitivity of the BIB argmax-persistence exponent to the
fitting convention (pooled $T_{\mathrm{argmax}}$, \BIB-\BIB, $m=50$,
$N=3$).  The first four columns are truncated-power-law fits at the
indicated $x_{\min}$, and the last is a pure power law at the Clauset
$x_{\min}$.  The bottom row is the cross-design spread
(max$-$min).}
\label{tab:xmin}
\begin{ruledtabular}
\begin{tabular}{lccccc}
Design & $x_{\min}$ auto & $x_{\min}{=}1$ & $x_{\min}{=}5$ & $x_{\min}{=}10$ & PL \\
\hline
rs & $1.441$ & $1.285$ & $1.441$ & $1.440$ & $1.599$ \\
ra & $1.428$ & $1.282$ & $1.428$ & $1.428$ & $1.578$ \\
ss & $1.438$ & $1.285$ & $1.438$ & $1.434$ & $1.598$ \\
sa & $1.409$ & $1.274$ & $1.409$ & $1.403$ & $1.574$ \\
\hline
spread & $0.032$ & $0.011$ & $0.032$ & $0.038$ & $0.024$ \\
\end{tabular}
\end{ruledtabular}
\end{table}

\subsection{Model selection for the persistence-time distributions}\label{app:modelselect}

Throughout, we report a power-law exponent only for conditions that pass a
fixed heavy-tail criterion.  For each pooled persistence-time set we select the
lower cutoff $x_{\min}$ by the Clauset--Shalizi--Newman
procedure~\cite{clauset2009} and compare five candidate distributions, namely
power law, truncated power law, exponential, lognormal, and stretched
exponential, by their Akaike weights on the common support $x\ge x_{\min}$.  A
condition is counted as heavy-tailed only if a power-law or truncated-power-law
form is Akaike-preferred over the exponential, the fitted range spans at least
two decades, and the gap between the pure- and truncated-power-law exponents is
at most $0.2$.

Under this criterion the \BIB-\BIB{} argmax-persistence and laminar-phase
distributions pass at every design in the core regime $\Nh\in\{6,10\}$ (argmax
spanning about three decades with a pure-versus-truncated gap near $0.16$,
laminar about three decades), as does the plateau observable (about $3.6$
decades, $\alpha\approx 1.23$).  The \BO-\BO{} distributions fail at every
design.  For random initialization the Akaike-preferred form is a stretched
exponential or a lognormal, and for structured initialization the best-fitting
truncated power law spans under $1.2$ decades with a pure-versus-truncated gap
above unity.  The same holds across the hypothesis-count sweep for all
$\Nh\ge 6$, and only at the degenerate $\Nh=|\mathcal{D}|=3$ do some \BO{}
designs pass.  \BIB{} itself fails the criterion at $\Nh\in\{15,20\}$, where the
surviving tail spans about one decade.  These are the finite-$\Nh$ boundaries of
the universal regime (Sec.~\ref{sec:fss-nhinv}), and no exponent is quoted
there.  The full per-condition verdict (winning model, Akaike weights, decade
span, and pure-versus-truncated gap) accompanies the deposited analysis code.

\subsection{Plateau-length distribution}\label{app:plateau}

The plateau observable of Sec.~\ref{sec:results-univ}, the maximal
argmax-stable intervals in which $P(h^{*})>\theta=0.3$ for a fraction $f=0.8$ of
the run, gives a stricter test of the collapse.  The \BIB-\BIB{} plateau-length
CCDF passes the heavy-tail criterion of Appendix~\ref{app:modelselect} (about
$3.6$ decades, gap $0.18$, $\alpha\approx 1.23$), whereas \BO-\BO{} does not.

\subsection{On-off intermittency of the log-posterior walk: supporting analysis}\label{app:heuristic}

The two-hypothesis reduction of Sec.~\ref{sec:soc-mech} places the
argmax persistence in the on-off intermittency class through the
first-return statistics of the log-posterior walk.  This appendix
collects the supporting analysis: the priority-queue analogy, the
verification of the drift identity Eq.~(\ref{eq:kldrift}), the
finite-drift scaling, and the boundary role of the conditional
smoothing.  The same first-return mechanism
underlies the heavy-tailed waiting times of human dynamics.  In the
priority-queue framework the $\tau^{-3/2}$ class arises at the
critical balance $\lambda=\mu$, where the queue length performs an
unbiased random walk and the waiting time is its first return to
zero~\cite{barabasi2005,vazquez2005} (reviewed
in~\cite{karsai2018}).  In that framework the exponent is not robust
to interactions.  Coupling two queues makes it drift with queue length,
from $2$ toward $1$~\cite{oliveira2009}.  \BIB{} instead reaches a
first-return exponent that is \emph{invariant} to its analogous
control parameter $\Nh$ (Sec.~\ref{sec:results-nh}).  Adding the
inverse-Bayesian relaxation step to inference yields the critical
regime intrinsically, with no control parameter to tune, so the
exponent is robust where the queue exponent is not.

The vanishing drift is the fixed point identified in
Sec.~\ref{sec:soc-mech}.  Across random $(p,L_i,L_j)$ triples the
measured per-step drift matches the identity Eq.~(\ref{eq:kldrift}) to
Monte-Carlo accuracy (Pearson $r=0.99999$), and in the production model
the within-phase drift falls below $10^{-4}$ once the renewal is active,
confirming that the inverse step pins the drift to zero structurally
rather than by tuning.

\begin{figure*}[tp]
\centering
\includegraphics[width=\linewidth]{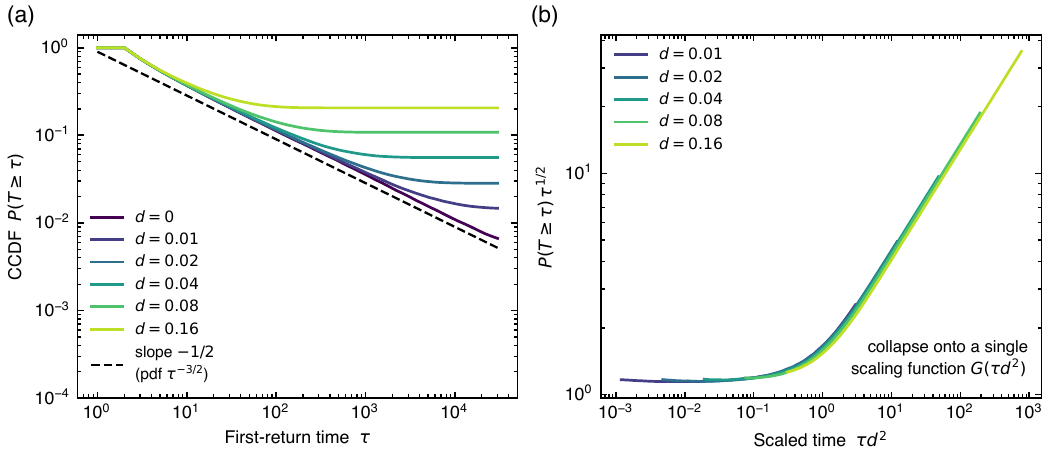}
\caption{The drift as the relevant variable, with $d=0$ the
critical line.  Reduced log-posterior walk $x_{t+1}=x_t+\eta_t$ with
tunable drift $d=\langle\eta\rangle$.  (a)~First-return CCDFs follow
$\tau^{-3/2}$ with a cutoff that retreats as $d\to0$.  (b)~Rescaling by
$\tau d^{2}$ collapses them onto a single scaling function $G(\tau
d^{2})$, confirming $P(\tau;d)=\tau^{-3/2}G(\tau d^{2})$ of
Eq.~(\ref{eq:driftscaling}).}
\label{fig:drift-scaling}
\end{figure*}

The first-return argument further predicts that the persistence
statistics are governed by the drift $d=\langle\eta\rangle$ itself, the
universal law being recovered only at $d=0$.  A reduced map
$x_{t+1}=x_t+\eta_t$ in which $\eta$ has tunable mean $d$ makes this
explicit.  Its first-return-time distribution obeys the finite-drift
scaling form
\begin{equation}
P(\tau; d) = \tau^{-3/2}\, G(\tau d^{2}),
\label{eq:driftscaling}
\end{equation}
with the Sparre--Andersen exponent $3/2$ on the $d=0$ line and a cutoff
$\tau^{\ast}\propto d^{-2}$, so the curves for different $d$ collapse
under $\tau\to\tau d^{2}$ (Fig.~\ref{fig:drift-scaling}).  The drift is
thus the relevant scaling variable and $d=0$ is the critical line.  By
the identity Eq.~(\ref{eq:kldrift}) the inverse-Bayesian renewal holds
the dynamics on this line as a fixed point of the inference rather than
a fitted condition, the operational sense in which the \BIB{} critical
state is tuning-free, arising structurally rather than by tuning.

The conditional smoothing enters only as a boundary operation.  For
$N=2$ the trigger $\min_i P(h_i)<0.002$ resets the losing mass from
below $0.002$ to $\approx 0.016$, i.e.\ it reflects $|u|$ from the wall
$u_{\mathrm{w}}=\log(0.998/0.002)\approx 6.21$ to
$u_{\mathrm{r}}=\log(0.984/0.016)\approx 4.12$.  This caps the longest
laminar phases, setting the truncated-power-law cutoff and its
$\Nh$-dependence (the $z$ exponent), while leaving the bulk walk,
and hence the $3/2$ law, intact.

Iterating Eq.~(\ref{eq:logodds-walk}) with this reinjection returns a power-law exponent
$\alpha_{\mathrm{PL}}\approx 1.50$ robustly across the increment scale,
confirming the $3/2$ first-return law, while the truncated-power-law
fit is pulled below $3/2$ (to $\approx 1.3$) by the cutoff.  The
measured exponents bracket $3/2$ in exactly this way, with the pooled
argmax data giving $\alpha_{\mathrm{PL}}\approx 1.59$ (above) and
$\alpha_{\mathrm{TPL}}\approx 1.43$ (below,
Table~\ref{tab:xmin}), consistent with a single $3/2$ on-off law read
through the two fitting conventions.  The same reduction reproduces
the smoothing ablation (Appendix~\ref{app:smoothing}).  Replacing the
conditional reinjection by an \emph{always-on} contraction
$u\mapsto\rho u$ ($\rho<1$) adds a per-step restoring drift, turning
the walk into an Ornstein--Uhlenbeck-like process with exponential
return times, so the heavy tail is destroyed.

We therefore assign the BIB persistence statistics to the on-off
intermittency class, with universal first-return exponent $3/2$
(equivalently, a $1/2$-stable limit for time-averaged observables in
the infinite-ergodic sense).  The reduction is a two-hypothesis,
independent-increment caricature.  The slight excess of the data
$\alpha_{\mathrm{PL}}$ above $3/2$ is attributed to the
multi-hypothesis competition at $\Nh>2$ and to correlations in the
per-step increment through the finite window $m$, a first-principles treatment of
which is left to future work.

\subsection{Matching-pennies control}\label{app:matching}

To test whether the BIB critical class is intrinsic to the
inverse-Bayesian rule rather than to the cyclic-dominance structure of
RPS (Sec.~\ref{sec:robust-mp}), we applied the identical inference core
to matching pennies, a two-action zero-sum game with a uniform Nash
equilibrium $(1/2,1/2)$ and no cyclic dominance.  Each agent keeps the
same $\Nh=10$ hypotheses, the same Bayesian update with conditional
Jelinek--Mercer smoothing, and the same inverse-Bayesian renewal (the
likelihood of the argmin-posterior hypothesis is replaced by the recent
$m=50$ empirical histogram), exactly as in RPS.  Only the observation
alphabet (two symbols) and the win/lose outcome change.  The
reward-based observation rule of Eq.~(\ref{eq:reward}) carries over
unchanged.  On a win the agent observes its own action, on a loss the
other symbol (matching pennies has no draw).  One agent is the matcher
(it wins when the two actions agree), the other the mismatcher.

Because the argmax-persistence exponent is design-independent in RPS,
and in particular invariant to the initialization (Fig.~\ref{fig:univ}),
we ran the two game-agnostic random-initialization designs (sampling
and argmax prediction), which transfer unchanged to a two-symbol
alphabet, varying only the prediction mode.  Each design was run for
$T=2\times10^{5}$ steps with a $10^{5}$-step burn-in and $20$
independent runs, pooling the argmax-persistence and laminar-phase
($\max_h P(h)>\theta$, $\theta=0.4$) run lengths over both agents and
fitting a truncated power law with the Clauset $x_{\min}$, as in the
main text.  Both observables reproduce the on-off $3/2$ class:
argmax-persistence $\alpha = 1.467$ and $1.471$, laminar-phase
$\alpha = 1.418$ and $1.419$ for the two designs, with the truncated
power law strongly preferred over an exponential in every case
(normalized log-likelihood ratio $R = 75$ to $89$ over
$n = 2.9\times10^{4}$ to $8.7\times10^{4}$ pooled events).  Both matching-pennies
exponents sit slightly above their RPS counterparts (argmax $1.47$
versus $1.43$, laminar $1.42$ versus $1.325$), the offset toward $3/2$
being consistent with the smaller finite-window drift at $N=2$
(Sec.~\ref{sec:robust-mp}).  The argmax value lies within the
structured-design band of the sharpness sweep ($1.44$--$1.51$,
Table~\ref{tab:sharpness}).  The
matching-pennies runs and fits re-use the production inference core
with $|\mathcal{D}|=2$.  The
cached run-length pools and the figure builder are included in the
deposited repository.

\begin{figure*}[tp]
\centering
\includegraphics[width=\linewidth]{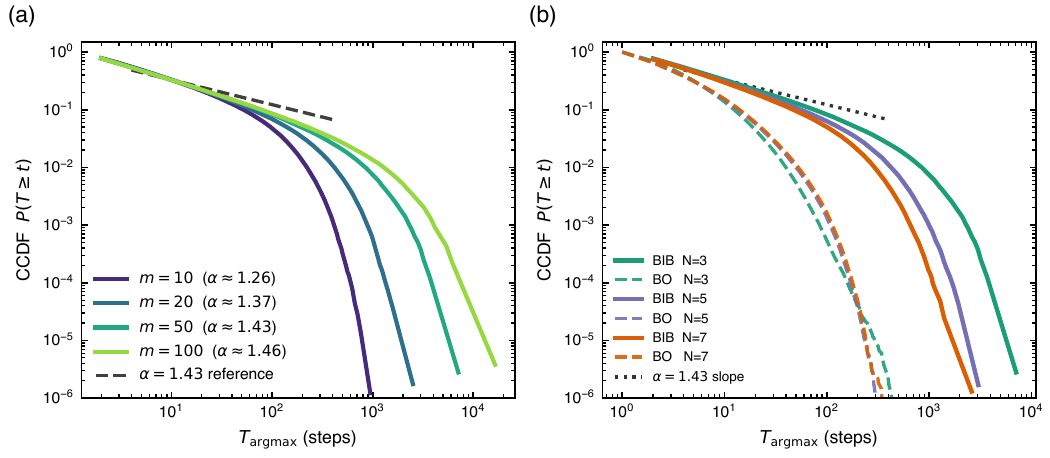}
\caption{BIB robustness sweeps along two control axes.
(a)~Window-size dependence, pooled \BIB-\BIB{} argmax-persistence
CCDFs $P(T\geq t)$ for window sizes $m\in\{10,20,50,100\}$ (pooled
across the four designs rs/ra/ss/sa).  At $m\le 20$ the tail is short
and the slope not yet converged.  By $m\ge 50$ the CCDFs settle onto the
canonical BIB power law, parallel to the $\alpha=1.43$ reference
(dashed, drawn over the power-law region and anchored on the $m=50$
curve).  (b)~Number-of-hands dependence, raw \BIB-\BIB{} (solid) and
\BO-\BO{} (dashed) argmax-persistence CCDFs at $m=50$ for
$N\in\{3,5,7\}$ (pooled over $4$ designs $\times\,2$ agents).  The BIB
tails keep a common heavy-tail slope close to the $\alpha=1.43$
reference (dotted) across all three $N$, whereas the BO tails decay
markedly faster and do not share a common slope.  As the BO
distributions do not collapse onto a single power law, no BO exponent
is fitted, and the distributions are shown directly.  $\Nh=10$ throughout.}
\label{fig:robust}
\end{figure*}

\subsection{Specificity among reinforcement-learning baselines}\label{app:rlbaselines}

To test whether the argmax-persistence ($\alpha\approx 1.43$) and
laminar ($\alpha\approx 1.325$) criticality is specific to \BIB{} or
generic to adaptive play on these uniform-Nash games
(Sec.~\ref{sec:learners}), we compared \BIB{} with three standard,
parameter-light learners under the main-text fit pipeline
(Appendix~\ref{app:xmin}): win-stay/lose-shift (WSLS)~\cite{nowak1993};
tabular $Q$-learning~\cite{watkins1992} (recall $1$, $\epsilon$-greedy
with $\epsilon=0.1$, learning rate $0.1$, discount $0.9$); and regret
matching~\cite{hartmascolell2000}.  Each was run against a
uniform-random opponent and in self-play for $T=2\times10^{5}$ steps
pooled over $40$ seeds, with the action encoding and payoff identical
to the \BIB{} agent (Sec.~\ref{sec:model-bib}).  For each agent we read
the persistence of its dominant internal preference (the argmax
hypothesis for \BIB{}, the argmax cumulative regret for regret
matching) or, for the reactive baselines, the played-action run
length.  Verdicts use a Clauset $x_{\min}$ with Akaike selection over
power-law, truncated-power-law, exponential, lognormal, and
stretched-exponential models, requiring a heavy-tail-preferred fit over
at least two decades with a pure-/truncated-fit gap $\le 0.2$.

Against a uniform-random opponent, regret matching is critical in the
same $3/2$ class as \BIB{}.  Because its cumulative regrets are
driftless random walks, the maximum-regret action persists as a
truncated power law with interior exponent $\alpha\approx 1.50$
(Sparre--Andersen, $3.39$ decades, gap $0.06$), while \BIB{} gives $\alpha\approx 1.45$
($2.97$ decades, gap $0.15$).  WSLS dwell times are geometric and
$Q$-learning's are better described by a lognormal than a power law.
The coincidence is imposed by the neutral environment, not
self-organized, and vanishes under self-play.  The regret walk acquires
a drift and its tail collapses to a lognormal, whereas \BIB{} retains
$\alpha\approx 1.43$ ($3.15$ decades, gap $0.16$).  Only \BIB{} is
critical across both opponent class and hypothesis count $\Nh$
(Sec.~\ref{sec:fss-collapse}): a meaningful $3/2$ claim requires both a
multi-decade power-law range and a small pure-/truncated-fit gap, which
\BIB{} satisfies ($\approx 3$ decades, gap $\lesssim 0.2$) and the
reactive baselines do not ($\approx 1$ decade, or a strongly curved
tail).

At the behavioral (played-action) level the picture is not diagnostic
of the inference rule (Table~\ref{tab:rlbehaviour}).  Against a fixed
biased opponent the action-run tail crosses over from exponential to a
heavy truncated power law for \BIB{} ($\alpha\approx 1.25$), but WSLS
reproduces the same crossover ($\alpha\approx 1.22$), while
$Q$-learning gives a stretched exponential and regret matching locks
onto the single best response.  The behavioral crossover is thus a
generic consequence of exploiting an exploitable opponent, consistent
with the caveat (Sec.~\ref{sec:scope}) that human-versus-bot signatures
are compatible with, but do not by themselves establish, \BIB{}.  The
discriminating signature is internal.  The baseline runs, fits, and
model-selection verdicts use $40$ seeds at $T=2\times10^{5}$.

\begin{table*}[tp]
\centering
\footnotesize
\caption{Behavioral (played-action) run-length class for each learner.
Against a neutral opponent the action stream is not a
learner-discriminator, and the crossover to a heavy tail against a biased
opponent is generic (WSLS matches \BIB{}).  Values are pooled over
$40$ seeds at $T=2\times10^{5}$ through the \BIB{} harness and the
main-text fit pipeline.}
\label{tab:rlbehaviour}
\begin{ruledtabular}
\begin{tabular}{lll}
Learner & Self-play & Vs.\ fixed biased $(0.6,0.2,0.2)$ \\
\hline
\BIB{}          & exponential (max ${\approx}13$)               & truncated PL, $\alpha\approx 1.25$ \\
WSLS            & degenerate ($2$-cycle)                        & truncated PL, $\alpha\approx 1.22$ \\
$Q$-learning    & non-critical (lognormal), $\alpha\approx 2.9$ & lognormal, $\alpha\approx 1.9$ \\
Regret matching & non-critical (lognormal), $\alpha\approx 2.6$ & locks onto one action \\
\end{tabular}
\end{ruledtabular}
\end{table*}

\subsection{Behavioral re-analysis of existing human play}\label{app:human}

This appendix gives the data and methods behind Sec.~\ref{sec:human}.
The behavioral corollary predicts that a player's behavioral
persistence is short-tailed against an adaptive opponent, which pins
the action stream toward the uniform Nash point, and crosses over to a
heavy tail against an exploitable opponent, onto which the agent locks.
We tested this on the public human-versus-bot data of Brockbank and
Vul~\cite{brockbank2024}, pooling their two experiments.  In
Experiment~1, participants played $300$ rounds against one of seven
bots with stable, stationary move patterns (hence exploitable by the
participant).  In Experiment~2, a separate participant pool faced
adaptive bots that instead exploited the participant's own
regularities.  We pool the fifteen bot conditions across both
experiments and order them by a single empirical axis, the human win
rate, which measures exploitability uniformly regardless of experiment.
Because the two experiments differ in design and recruited separate
participants, the gradient is a cross-condition trend, not a
within-subject manipulation.  After the $\ge 50$-round inclusion filter
the pool comprises $n=451$ completed games ($244$ against the fixed
bots, $207$ against the adaptive bots).  As these bots are
transition-based, the strategically relevant persistence is in the
player's transition sequence, which we therefore analyze.

Both behavioral predictions hold (Fig.~\ref{fig:human}).  The
transition-run tail is markedly heavier against the fixed than against
the adaptive bots, and across the fifteen conditions the mean
transition-run length increases monotonically with empirical
exploitability (Spearman $\rho=+0.85$, $p=10^{-4}$), with a truncated
power law preferred over an exponential in every condition.  This is
consistent with, but does not by itself establish, the internal
criticality of Sec.~\ref{sec:soc}, since the argmax-persistence
statistic is a latent property of the inference and is not directly
observable in the human symbol stream.  The re-analysis uses the
authors' public dataset (\texttt{github.com/erik-brockbank/rps}).  The figure and the
$\rho=+0.854$ gradient are reproduced from a cached condition summary.

\subsection{Distinction from standard resampling}\label{app:resampling}

A natural question is whether the inverse step is simply a standard
remedy for posterior degeneracy (Sec.~\ref{sec:soc-mech}).  We compared
it, on the same substrate (\BIB-\BIB{}, rs, $\Nh=10$, $m=50$,
$T=2\times10^{5}$), with sequential importance resampling (SIR), the
bootstrap particle filter that resamples the hypotheses in proportion
to their posterior when the effective sample size drops, with light
jitter~\cite{gordon1993}, and with a prior-restart that reinitializes
the least-supported hypothesis from the prior at the \BIB{} cadence.
The argmax-persistence tail is heavy for all three, so that observable
alone is not diagnostic.  The discriminating signature is the on-off
laminar structure.  Only the directed
renewal sustains intermittent single-hypothesis dominance, with long
laminar phases (mean $\approx 41$ steps, $\alpha^{\mathrm{lam}}=1.325$).
Prior-restart produces only short dominance episodes (mean $\approx 6$
steps), and SIR sustains \emph{no} laminar phases at all, the posterior
staying diffuse as under Bayes-only.  The renewal from the recent
empirical pattern is what lets a hypothesis briefly dominate yet be
overtaken, the mechanism of the on-off alternation, rather than
either never dominating (SIR) or being disrupted at once
(prior-restart).  The comparison uses $40$ seeds at $T=2\times10^{5}$.

% =============================================================================
\subsection{Robustness-sweep figures}\label{app:robustfigs}

\begin{figure*}[tp]
\centering
\includegraphics[width=\linewidth]{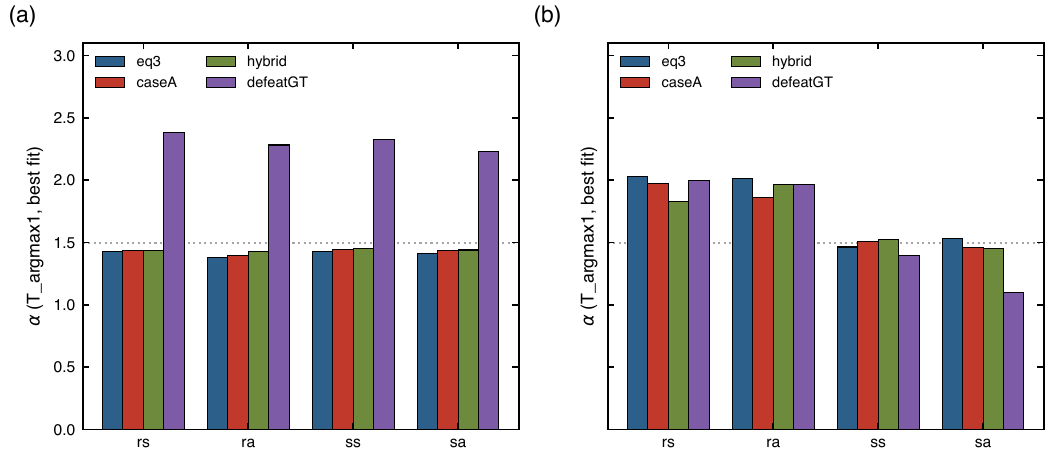}
\caption{Observation-rule ablation at medium scale
($T=10^{4}\times\,200$ runs).  (a,~left)~\BIB-\BIB{}
argmax-persistence exponent $\alpha(\Targmax)$ by design and scheme.
The three renewal-preserving schemes (eq3, caseA, hybrid) cluster tightly at
$\alpha\approx 1.4$ while defeatGT lands on a different attractor at
$\alpha\approx 2.3$.  (b,~right)~same for \BO-\BO{}, included for
completeness.}
\label{fig:scheme-ablation}
\end{figure*}

The observation-rule ablation of Sec.~\ref{sec:robust-obs} is collected
here (Fig.~\ref{fig:scheme-ablation}), and the window-size and
game-dimension sweep of Sec.~\ref{sec:robust-mN} is shown in
Fig.~\ref{fig:robust}.

\bibliographystyle{apsrev4-2}
\bibliography{refs}

\end{document}